\documentclass[11pt,a4paper]{article}
\usepackage{epsfig,graphics,rotating,color}
\begin{document}
%\begin{titlepage}
\title{Are multimodal events a sign of the strangelet passage through the matter?}
\author{Ewa G\l{}adysz-Dziadu\'s\\
e-mail: edziadus@gmail.com\\
Institute of Nuclear Physics Polish Academy of Sciences, Cracow, Poland \footnote{retired}}
\maketitle
%\begin{abstract}
We investigate the possible connection between the multimodal events (MME) observed   
in very high energy extensive air showers 
by the HORIZON experiment
and the so-called strongly penetrating component observed in the homogenous lead emulsion chambers 
of the Pamir and
Chacaltaya Experiments. We found that both experimental observations 
 could be connected one to the other, and could be 
the manifestation of the same physical process, i.e. penetration of a strangelet through 
the matter. In the first case a strangelet produces the many-maxima long range
cascades observed in the homogenous lead emulsion chambers. In the second one the successive interactions
of a strangelet in the air are seen in the HORIZON detectors as the consecutive signals. Time intervals
between signals are between several dozen to  several hundred nanoseconds. 
%\end{abstract}
%\end{titlepage}
\section
{\Large Introduction}

The characteristics of very high energy cosmic-ray interactions 
over $\sim 10^{15}$ eV reveal some anomalies 
which  are hardly  explained by the physical rules known from
the  hadronic interactions in lower energy accelerator experiments.
 The wide spectrum of unusual events, such as for example Centauro species
 \cite{1,2,3},  the so-called strongly penetrating component appearing
 in the form of long-range many-maxima cascades  
\cite{5,6}, multimodal events (MME) \cite{Beisembaev_5peaks, Beznosko_unusual_events, peaks-3-8-4-6},
 was observed above this threshold.  
Monte Carlo simulations show that 
 any kind of statistical fluctuations of "normal" hadronic interactions could 
not produce such anomalies.  

In the past years some explanations based on the idea of the QGP and/or strangelet formation 
have been proposed for Centauros  \cite{7,8,9} and for the 
strongly penetrating component \cite{strangelet_Pb}).

In particular in the paper \cite{strangelet_Pb}, entitled {\it Is the strongly penetrating
component a sign of the strangelet passage through the matter?} we showed that {\it "the many
maxima cascades observed in the thick homogenous lead emulsion chambers \cite{5,6}  of the Pamir and 
Chacaltaya experiments can be produced in the process of a strangelet penetration through the
apparatus"}. 
For description and proposed interpretations of these unusual phenomena 
see the reviews \cite{Ewa_review, Kempa_review}.

In this work we attempt to estimate how could manifest the strangelet  passage through the air 
and to consider the question of their possible existence  
 in the present experimental data. Our considerations indicate 
that the puzzling observations of the HORIZON 
experiment,
could be explained by the passage of strangelets through the air.
It seems that the mechanism which we proposed and used to explain the 
many maxima strongly penetrating cascades 
observed in the thick Pb chambers of the Pamir Experiment \cite{strangelet_Pb} 
could be also responsible for the multimodal events
found in the HORIZON detectors.  

\section
{\Large The experimental puzzles}

\subsection{The strongly penetrating many-maxima cascades}
The strongly penetrating component is one of the most surprising phenomena  
observed in the mountain emulsion chamber experiments \cite{5,6, strangelet_Pb, Ewa_review}.
 This phenomenon manifests itself  by the
characteristic
energy pattern revealed in the shower development in the deep chambers
(calorimeters) indicating
the slow attenuation and many maxima structures.

 The very spectacular examples of such phenomena are exotic cascades which we have found in the
Centauro--like event C--K \cite{5} detected in the deep homogenous type
  Pb chamber of the Pamir experiment. It was exposed at the altitude of 
4370 m a.s.l. 
The thickness of the chamber was 60 cm of the lead, what corresponds to $\sim$ 3.5 interaction lengths
for hadrons or $\sim$ 108 cascade unites. The cascades were detected at the X-ray films.  
 The event, similarly as other 
Centauro species is the hadron-rich family, consisting of 123 cascades. The energy deposited by the 
photon cascades and hadron cascades was $\Sigma E_{\gamma}$ = 306 TeV and  $\Sigma E_{h}^{\gamma}$ = 392 TeV
respectively. So, the total energy of the event was $\sim 10^{15}$ eV. The zenithal angle of the 
event, calculated as the average zenithal angle
of all registered cascades is $<\theta> = 27.4^{0}$. Among them 
three
 hadron cascades,  emitted in the very forward rapidity region, 
seem to be not "ordinary" ones and their transition curves  exhibit surprising 
features:

\begin{itemize}
\item strongly penetrative nature associated with a very slow attenuation
\item appearance of many maxima (11, 5 and 3 humps respectively)
 satisfactorily fitted by the individual electromagnetic cascade curves
\item small distances between the maxima, being about two times shorter 
than the calculated distance for the
"normal" hadron cascades \cite{Iwanienko})
\end{itemize}  

\begin{figure}[h]
\begin{center}
\includegraphics[width=1.\linewidth]{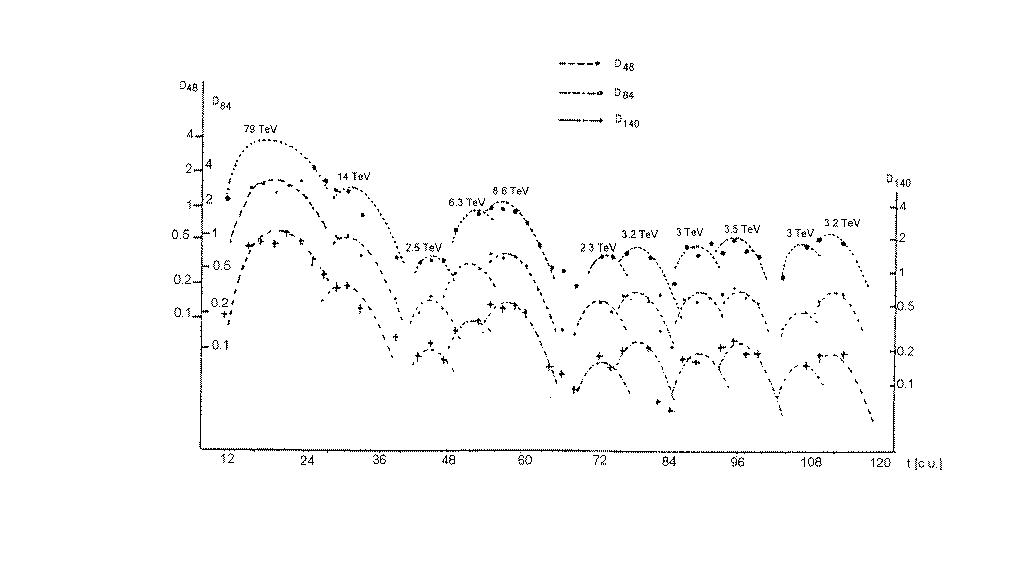}
\vspace{-1.7cm}
\caption{
 Transition curves 
 in X-ray film darkness D (measured
in
three diaphragms of a radius R = 48, 84 and 140 $\mu$) for cascade
no 197.08. Energy (in TeV units) liberated into the
soft component is indicated at each hump \cite{5}.}
\vspace*{-0.5cm}
\label{fig:cascade}
\end{center}
\end{figure}

 Transition curves of the longest 
cascade is shown in Fig.~\ref{fig:cascade}.

  This exotic cascade  penetrated more
than 109 cascade units and 11 maxima appeared along  its transition
curve. The average distance between the neighbouring humps of many-maxima cascades detected  
in the "Pamir 76/77" chamber \cite{5} has been found  to be  7.2 cm Pb (i.e. $\sim 0.4 \lambda_{int}$
for hadrons in Pb).
  
The other examples of such anomalous cascades, found in the deep Pb chambers have been  
 described    in \cite{6}.
Simulations,  assuming 
standard  models
of hadron-nucleus interactions
  confirmed the unusual character
of these long-penetrating cascades \cite{Tamada}.

\subsection{Multimodal events}

HORIZON experiments are aimed to study spatial and
temporal distributions of the Extensive Air Showers (EAS) of the energy above $\sim 10^{16}$ eV.
These detector systems are located at the  Tien Shan Science Station,
at the altitude $\sim$ 3340 meters
above the sea level \cite{Horizon-8T, Horizon-10T}.

A sizable number of  events detected during all runs of the Horizon experiments have features different
from the "standard" events. They revealed  many maxima structures seen  in
several detection stations of the Horizon system
\cite{peaks-3-8-4-6}. 
These events called {\bf MME}, i.e. {\bf multimodal events}  
 exhibit the unusual spatial and temporal 
structure of pulses: 

\begin{itemize}
\item  Pulses consist of several smaller peaks (called also modes or maxima) and are {\bf separated one to 
the other by  distances from few tens to hundreds of nanoseconds.} 
\item  Separation between the maxima 
increases with a distance of detectors from the EAS
core.
\end{itemize}

 This phenomenon can not be obtained from simulations based on the standard models.

Below are shown two examples of the typical unusual events that follow a multipeak behaviour. 

\begin{enumerate}

\item  {\bf 5-peak MME event \cite{Beisembaev_5peaks} registered by the Horizon-10T system on April 6, 2019.}

The event is shown in  Fig.~\ref{fig:MME_5peaks}.

\begin{figure}[h]
\begin{center}
\includegraphics[width=0.8\linewidth]{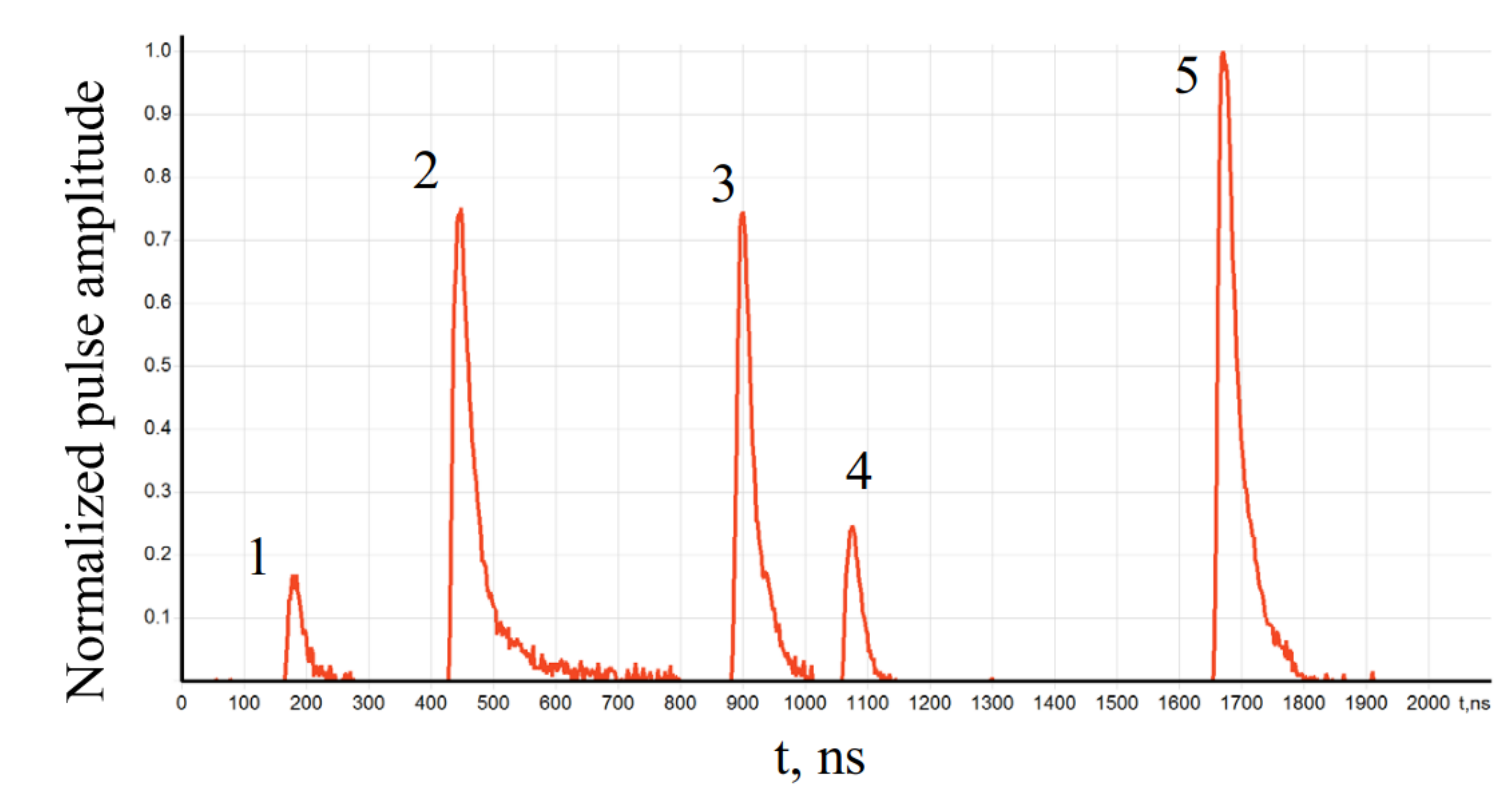}
\caption{Example of the 5-peak multimodal event detected 
at $R \approx 898$ m from the axis of the event by the Horizon-10T system on April 6, 2019,
 reprinted from \cite{Beisembaev_5peaks}.}
\label{fig:MME_5peaks}
\end{center}
\vspace*{-0.4cm}
\end{figure}

  The EAS axis of the event
arrived from the Southern direction with the zenithal angle $\theta = 52^{0}$
and it was registered near the detection
point  no 7 and closer to the point no 6.  Fig.~\ref{fig:MME_5peaks} shows pulses registered
by 1 m$^{2}$ scintilator  located at the detection point no 9, positioned at the distance 
 R $\approx$ 898 m from the 
EAS axis. 
Five separate pulses can be clearly distinguished.
 Tab.~\ref{tab-5peaks} shows their main characteristics: the delay of each pulse 
from the first one \cite{Beisembaev_5peaks} and 
the resulting time distances
$\Delta\tau$  between
the successive peaks.

\begin{table}
\begin{center}
\begin{tabular}{|c|c|c|c|c|c|}
\hline
\multicolumn{6}{|c|}{}\\
\multicolumn{6}{|c|}{Characteristics of 5-peak MME event}\\
\multicolumn{6}{|c|}{}\\
\hline
Peak no & 1 & 2 & 3 & 4 & 5\\
\hline
Delay [nsec] & 0 & 265 & 720 & 894 & 1493\\
$\sim \Delta\tau$ [nsec] & - & 265 & 455 & 174 & 599\\
\hline
 \end{tabular}
  \end{center}
\caption{
Characteristics of the MME event from April $6^{th}$, 2018, Horizon-10T \cite{Beisembaev_5peaks}.}
\label{tab-5peaks}
\end{table}

According to \cite{Beisembaev_5peaks}, the characteristics
of this event  are totally different than these observed as well as  
in "usual" experimental and simulated by the CORSIKA package events.
In particular:
\begin{itemize}    
  \item The simulated EAS  produces a single pulse
 with a uniform structure without breaks.  It is in contrary 
to the multipeak structure of signals in this event.
 \item The width of the pulse at R $\approx$ 898 m should be
 $\sim$ 600 ns. The measured widths of peaks are between 28 ns and 35 ns \cite{Beisembaev_5peaks}.
\item  The disk thickness increases and 
 the particle density in the disk decreases with a distance from the EAS axis.
At R $\approx$ 898 m from axis, the average particle density for simulated events of energy
$E_{0} = 10^{18}$ eV is below one particle/m$^{2}$ \cite{Beisembaev_5peaks}.
 It is in contrary to  $\sim$ 173 particles/m$^{2}$ found in this event. 
\end{itemize}

In conclusion, the event with such characteristics can not be described
by the standard  models used for description of the EAS.

\item {\bf The MME event \cite{peaks-3-8-4-6} detected by the Horizon-10T detector system 
\cite{Horizon-10T} on Jan. 2016}
 
This system consists of 10 detection stations which are located at different distances from the 
central station named "Center". Their names and numbers, and distances R[m] from the central
detection point are published in \cite{Horizon-10T}. The distance of the farest station, named "Bunker",
 from the central detection point 1 is 1000 m. 

The event is shown in Fig.~\ref{fig:MME-4-6}.
Analysis of signals \cite{peaks-3-8-4-6} indicated the following features of the event:
\begin{itemize}
\item The very large signal
has been detected at the point 2 and the next in time signal has been seen by detectors of the station 
no 5. Peak structures were not observed in these points. It proves  that EAS axis has arrived
somewhere between these points, closer to the point no 2.
  
\item The multimodal structures have been observed in 
 detectors of all remaining stations. 
The character of the observed structures strongly depends    
on the distance $\Delta R_{ia}$ of the detection point $i$
from the EAS axis $a$.
 
\item The pulses observed in the detection points 1 and 7, 
located approximately at the same distance from the EAS axis and further than the point 5, have complicated 
structures. However, because of rather small  distances of points 1 and 7 from the EAS axis 
 it was difficult to measure the time intervals between the particular modes \cite{peaks-3-8-4-6}. 
   
\item The pulses from detection points 6 and 4 demonstrate apparent multimodality. Peaks are
separated by hundreds of nanosecs, see Fig.~\ref{fig:MME-4-6}.
As it is seen in the Table ~\ref{tab:delta_tau} these points are located far from the point no 2, 
assumed to be  the EAS axis.
However, some of the pulses have
a complex shape, what could indicate that they possibly consist of more than one signal. In a consequence   
 the true values of  $\Delta\tau$ are a little shorter \cite{peaks-3-8-4-6}. 

\item The apparent peak structures are observed also at the detection points 8 and 3, see the right 
picture of Fig.~\ref{fig:MME-4-6}.
 They  are located far from the EAS axis and 
 the successive modes  are separated by several hundreds of nanosecs. As the station
no 3 is located at the furthest point from the EAS axis, it is possible that some modes could be 
 too weak to be detected. In a consequence 
only 2 peaks were observed.  

\end{itemize}

So, in our further considerations, we assume that the center
of the EAS was approximately  in the point no 2. Taking from \cite{ peaks-3-8-4-6, Horizon-10T} coordinates 
X, Y, Z of all stations we have calculated the distances $\Delta R_{ia}$ of the detection point $i$
from the EAS axis $a$ (see Table~\ref{tab:delta_tau}).

The values of time distances  $\Delta \tau$ between the consecutive modes, estimated from Fig.~\ref{fig:MME-4-6}
  are shown in the Table~\ref{tab:delta_tau}.
It is seen  that the signals detected by stations located further from
the event axis are more delayed. The correlation between 
the length of time intervals  $\Delta \tau$ and the time arrivals of the successive signals is also observed 
by each station.  
{\bf More delayed signals have longer $\Delta \tau$.}  

As it will be described in the next
section, such  features  are in agreement with the proposed by us explanation of the MME events.

\begin{figure}
\vspace{-1cm}
\begin{center}
\includegraphics[width=0.49\linewidth]{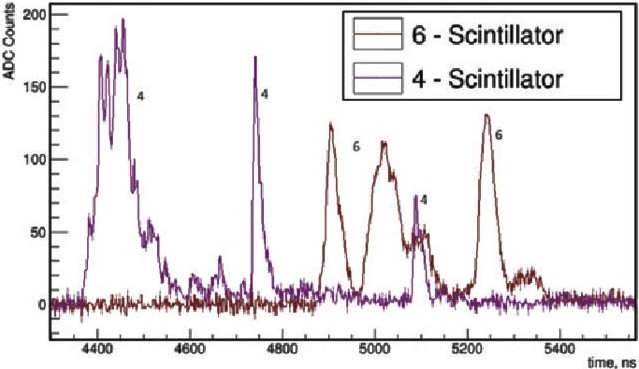}
\includegraphics[width=0.49\linewidth]{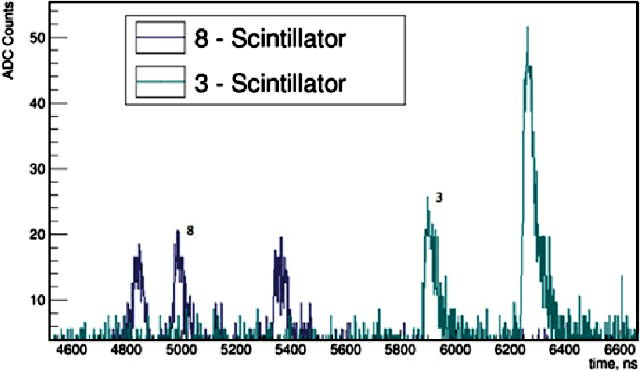}
\caption{The MME event seen by the HT detector, reprinted from \cite{peaks-3-8-4-6}.
 3 peaks in the station no 8 and 2 peaks
in the station no 3 have been detected (right picture).
 3 peaks have been detected as well 
as in the station no 4 and  in the station no 6 (left picture).}
\label{fig:MME-4-6}
\end{center}
\vspace*{-0.5cm}
\end{figure}

\begin{table}
%\vspace*{-1.5cm}
\begin{center}
\begin{tabular}{|c|c|c|c|}
\hline
\multicolumn{4}{|c|}{}\\
\multicolumn{4}{|c|}{ Time delay $\tau$ of signals and time distances $\sim \Delta \tau$ between them}\\
\multicolumn{4}{|c|}{}\\
\hline
Station no & $R_{2i}$ [m] & Peak delay $\tau$ [nsec] & $\sim \Delta \tau$ [nsec] \\
\hline
 4 & 392 & 4375      & 0  \\
 4 & 392 & 4750      & 375\\
 4 & 392 & 5100      & 350  \\
 6 & 601 & 4900  & 0 \\
 6 & 601 & 5050  & 150 \\
 6 & 601 & 5250 & 200 \\
 8 & 769 & 4800 & 0 \\
 8 & 769 & 5000 & 200 \\
 8 & 769 & 5350 & 350 \\
 3 & 848 & 5900 & 0 \\
 3 & 848 & 6300 & 400\\
\hline
 \end{tabular}
  \end{center}
\caption{Delay of signals $\sim\tau$ [nsec]  and time distances between the peaks 
$\sim \Delta \tau$ [nsec]
for detection points 4, 6, 8 and 3 in the MME event
detected on Jan.26 2016 by the HT experiment. Calculated distances of the station $i$ from 
the event axis, being estimated 
to be close to the point 2, $R_{2i}$  are shown.}
\label{tab:delta_tau}
\vspace*{-0.5cm}
\end{table}

\end{enumerate}

In general all MME events found by the Horizon detectors follow the same pattern:
\begin{itemize}
\item  the large single 
pulse (that may have a peak structure too
fine to be resolved) seen close to the center of the event
\item  the complicated structures seen at the intermediate
distances from the axis
\item   apparently separated peaks at the furthest points from the EAS axis
\item the longest
time distances $\Delta \tau$ between the peaks  observed for the most delayed pulses
\end{itemize}

  Summarizing, {\bf the main
feature of the unusual MME events is the increase of the time intervals  between the successive peaks 
with both the distance of the detection station from the center of the event (event axis) 
and the time delay of the pulse relative to the first in time pulse registered by this station.}

As it will be explained  in the next section, we  assume that the earliest in time peak
 comes from the highest altitude. It is possibly produced by the EAS particles born 
in the first strangelet interaction  with the air nuclei.
In such scenario as well as the delay of the successive peaks and the length of  time intervals between
them decreases with the increase of the altitude of a possible strangelet interaction. On the other 
hand the length of the time distances $\Delta\tau$ between the successive modes  is increasing with 
the distance of the 
detection point from the EAS axis. All these predicted by our scenario features are observed not only 
in the above described 
events but they are also   
in accordance with the experimentally observed tendency of other MME events.   

\section{\Large Proposed explanation}

We propose and investigate the hypothesis that the mentioned in the previous section phenomena
are closely related one to the other and the same mechanism could be responsible for 
their appearance. Our considerations indicate that both phenomena could be 
 manifestation of a strangelet passage through the matter. 
As well as the strongly penetrating many-maxima cascades observed in the thick Pb chambers 
and the multimodal events could be produced in the process of the successive strangelet interactions 
 with the lead nuclei in the first case or with the air nuclei in the second one.
      
\subsection
{Many-maxima cascades as signs of a strangelet passage through the thick lead chambers}
 
Strangelets are hypothetical droplets of the stable strange quark matter \cite{Bodmer, Terazawa, Witten}.
 To answer the question of whether a strangelet
could be detectable in the typical emulsion chambers used in the mountain 
experiments at Mt. Chacaltaya and Pamirs and what should be signs of its passage 
through the apparatus we have considered the 
several possible decay \cite{Berger, Chin, Greiner} and interaction \cite{Wilk}
modes of  strangelets and we have simulated their travelling through the
homogenous-type thick lead chambers or tungsten calorimeters.
 The method and our results are described in \cite{strangelet_Pb, EPJ_gladysz}.
Our simulations  showed \cite{strangelet_Pb} that  
{\em strangelets could  produce in the thick chambers/calorimeters 
with the heavy absorbers  the long many-maxima
transition curves resembling the observed long-lived cascades}.

 As we have mentioned in 
\cite{EPJ_gladysz} the strong
penetration power of some objects observed in the cosmic ray experiments
can be connected  with both the small interaction cross section of a
strangelet and  a  big concentration of its energy in a narrow 
region of a phase space. This energy  could be  liberated into
a conventional particle production in many consecutive interaction (or evaporation) acts.
 As the result  we proposed {\em the energy deposition pattern in  deep calorimeters
as {\bf  a new 
strangelet signature \cite{EPJ_gladysz}}}.  
In our opinion {\em this idea could  be extended and applied also for interpretation
of the surprising results of the HORIZON experiments}. We suppose that {\em {\bf
 the consecutive interactions of 
a "stable" strangelet with the air nuclei could generate the multimodal events, observed in the form 
of separated in time signals.}}
      
 We call {\bf "stable" strangelets}  \cite{strangelet_Pb, EPJ_gladysz} the long-lived objects capable to 
reach and pass through the apparatus without decay. 
As it has been estimated in \cite{Chin}  the hyperstrange multiquark
droplets, having the strange to baryon ratio $f_{s}$ = 2.2 - 2.6 can be the 
subject of only the weak leptonic decays and their lifetime is estimated
 to be longer 
than $\tau_{0} \sim 10^{-4}$ sec.
 Such long-lived objects 
 should pass without decay not only through the thick Pb chambers 
but also, before evaporation into the bundle of neutrons, could travel 
the long distances 
  in the atmosphere.
While traveling, the "stable" strangelet collides and interacts with the matter.
Following the phenomenological indications of \cite {Wilk} we have simulated and described 
 their possible interactions in the
lead chambers  \cite{strangelet_Pb} and in the thick tungsten calorimeters \cite{EPJ_gladysz}.

 The strangelet was considered as the object of the radius
\begin{equation}
R_{str} = r_{0} A^{1/3}_{str}
\end{equation}
where the rescaled radius
\begin{equation}
r_{0} = ({\frac{3 \pi}{2(1 - \frac{2 \alpha_{s}}{\pi})[\mu^{3} + 
(\mu^{2} - m_{s}^{2})^{3/2}]}})^{1/3}
\end{equation}
$\mu$ and $m_{s}$ are the chemical potential and the mass of the strange quark 
respectively and $\alpha_{s}$ is the QCD coupling constant.
 The mean 
interaction path of strangelets in the lead absorber
\begin{equation}
\lambda_{s-Pb} = \frac{A_{Pb} \cdot m_{N}}{\pi(1.12 A^{1/3}_{Pb} + 
r_{0}A^{1/3}_{str})^{2}}
\end{equation}

Penetrating through the Pb chamber a strangelet collides with the lead nuclei. In
each act of a collision, the "spectator" part of a strangelet
survives (a strangelet is much more strongly bound than a normal
nucleus) continuing the passage through the chamber. The "wounded"
part is destroyed.
 The process is ended, when in the successive
interaction the whole strangelet is destroyed. Particles, generated
in  the 
consecutive collision points, interact  with the absorber nuclei in a "normal" way,
resulting in the electromagnetic-nuclear cascade, developing in the 
chamber matter.

The general conclusion from our studies was that "stable" strangelets  
 can produce in the thick Pb chambers or tungsten calorimeters long range many
maxima cascades. The distances between the successive  peaks depend  both on 
the value of the mean interaction path of the strangelet in the absorber matter $\lambda_{str}$ 
and on the value of
the mean interaction paths of usual particles.
  
As we will show in the next sections the puzzling results of the HORIZON experiments also support 
 the picture of a "stable" strangelet passage through the air layer
above the EAS detectors.

\section{Are many-maxima MME  events observed by Horizon experiments signs of a strangelet
passage through the air?}

The possibility of appearance of many maxima structures and their character depends 
on several factors. The most important are both  the absorber properties 
and the type of interacting objects.  
In particular  the  mean interaction path of objects interacting with the absorber nuclei 
 should be  long enough to resolve the successive interactions. From the other hand  it must be
 short enough to give opportunity for several projectile interactions in the absorber 
layer. It is the reason that the 
direct comparison of many maxima structures observed in the Pamir and in the Horizon experiments is
not possible. In the first case we deal with the heavy absorber (Pb) and in the second one with
the light nuclei of the air. So, at first we should look at and compare the values of the
mean interaction paths of interacting objects in both kind of the experiments.

\subsection{Strangelet mean interaction path in the Pb absorber}

It is obvious that the  mean distances between peaks, observed in the Pamir Pb chambers, should depend on
the interaction length $\lambda_{str}$ of strangelets passing through the absorber.   
 
\begin{figure}[t]
\vspace{-0.5cm}
\begin{center}
\hspace*{-1cm}
\includegraphics[width=0.52\linewidth]{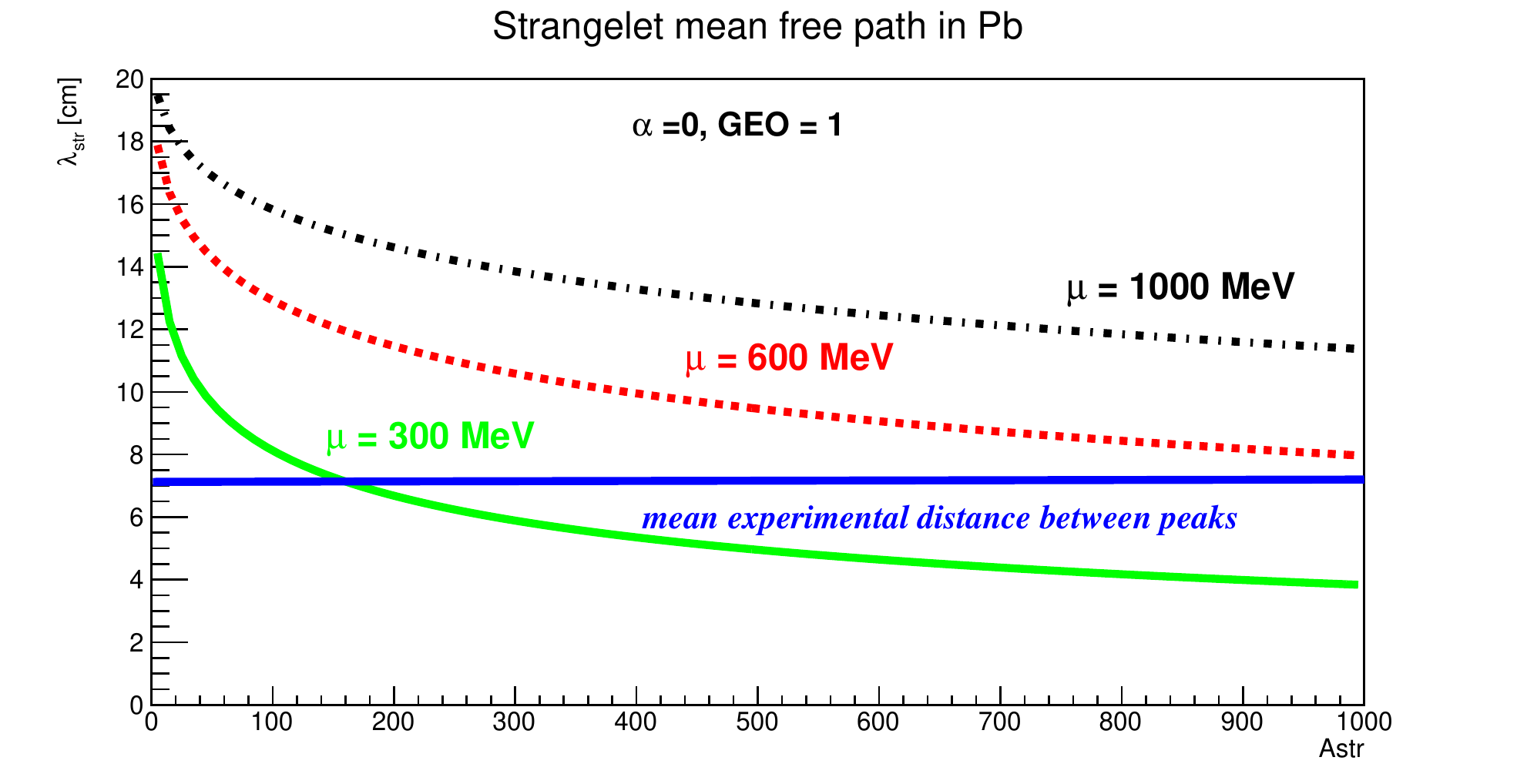}
\includegraphics[width=0.52\linewidth]{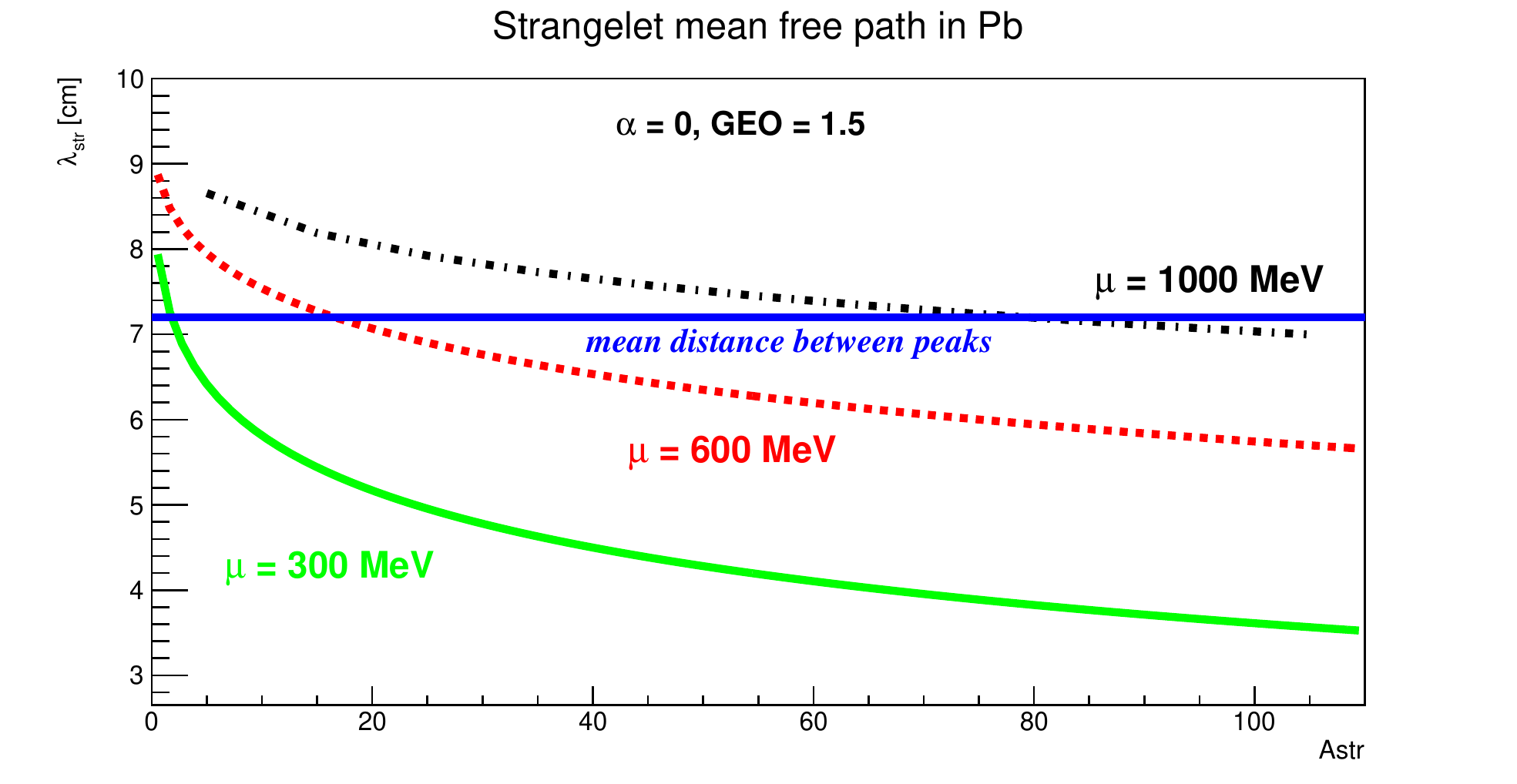}
\caption{Calculated strangelet mean interaction path $\lambda_{str}$ [cm] in Pb absorber 
as a function of a strangelet mass number $A_{str}$,
for $\alpha_{s}$ = 0 and GEO = 1 (left picture) and GEO = 1.5 (right picture). 
The mean distance between experimentally
observed peaks \cite{5} is shown by the blue line.} 
\label{fig:Pb_alfa0}
\end{center}
\vspace*{-0.8cm}
\end{figure}

Fig.~\ref{fig:Pb_alfa0}
shows $\lambda_{str}$  
in the Pb absorber as a function of a strangelet mass number $A_{str}$.  These calculations have been performed
for  $\alpha_{s}$ = 0,
 for two different values of the geometrical factor (GEO = 0 or GEO = 1.5) and for three values of a strangelet 
quark-chemical potential ($\mu$ = 300, 600 and 1000 MeV). We used the 
procedure and equations described in the section 3.1.
The curves obtained for  $\alpha_{s}$ = 0.3 are very close to these 
shown in Fig.~\ref{fig:Pb_alfa0}, so they are not presented here.
The mean values of the experimentally observed  distances between the consecutive peaks 
of many maxima cascades \cite{5} are marked
 by the blue straight lines.  
As it is illustrated in Fig.~\ref{fig:Pb_alfa0}
the blue lines  lie 
 between the calculated  $\lambda_{str}$ curves.
It indicates  the close relations between the calculated values of  $\lambda_{str}$ and
the measured distances between the successive peaks of the Pamir  many-maxima cascades.
As it was mentioned in the previous section, this statement has been confirmed in our simulations. 
Simulating the passage of strangelets through
the thick Pb chambers we have obtained the long many- maxima cascades resembling the experimentally found
unusual events \cite{strangelet_Pb}. 

The observed many maxima structures
could arise in the result of 
a passage of rather low mass ($Astr < 100$) strangelets through the thick Pb chambers 
\cite{strangelet_Pb, EPJ_gladysz}.    
 The high mass strangelets could be rather responsible for 
the phenomenon called "halo". These very high energy events have been  observed at the X-ray 
films in the form of large area black spots,
 penetrating through the whole chamber. In this case it was impossible to separate 
the individual peaks. For description of various forms of the observed strongly penetrating component
see our review \cite{Ewa_review}. 

So, {\bf the assumed scenario 
seems to be  the good direction for understanding 
of the origin and the properties of the unusual many-maxima events}.
Complementing the conclusions of our earlier works \cite{strangelet_Pb, EPJ_gladysz}
we can say:
\begin{itemize}
\item The calculated values of a strangelet mean interaction paths $\lambda_{str}$ in Pb turned out to 
be appropriate for explanation
of the observed values of distances between the successive peaks of many-maxima cascades. 
\item  The calculated  distances between the successive peaks
  produced 
by "standard" particles  are about two times longer \cite{Iwanienko}. They would be located 
not only far from the experimental blue line but also far from  the calculated $\lambda_{str}$
values. 
\end{itemize}

\subsection{Strangelet mean interaction path in the air}

Similarly as  in the case of the thick Pb chambers, 
the idea of a possible strangelet detection by the EAS stations is based on the expectation that strangelets 
should have geometrical radii much smaller than the ordinary nuclei of the same mass number,
 and correspondingly,
much smaller interaction cross sections. 

\begin{figure}
\vspace*{-1.5cm}
\begin{center}
\hspace*{-1cm}
\includegraphics[width=0.52\linewidth]{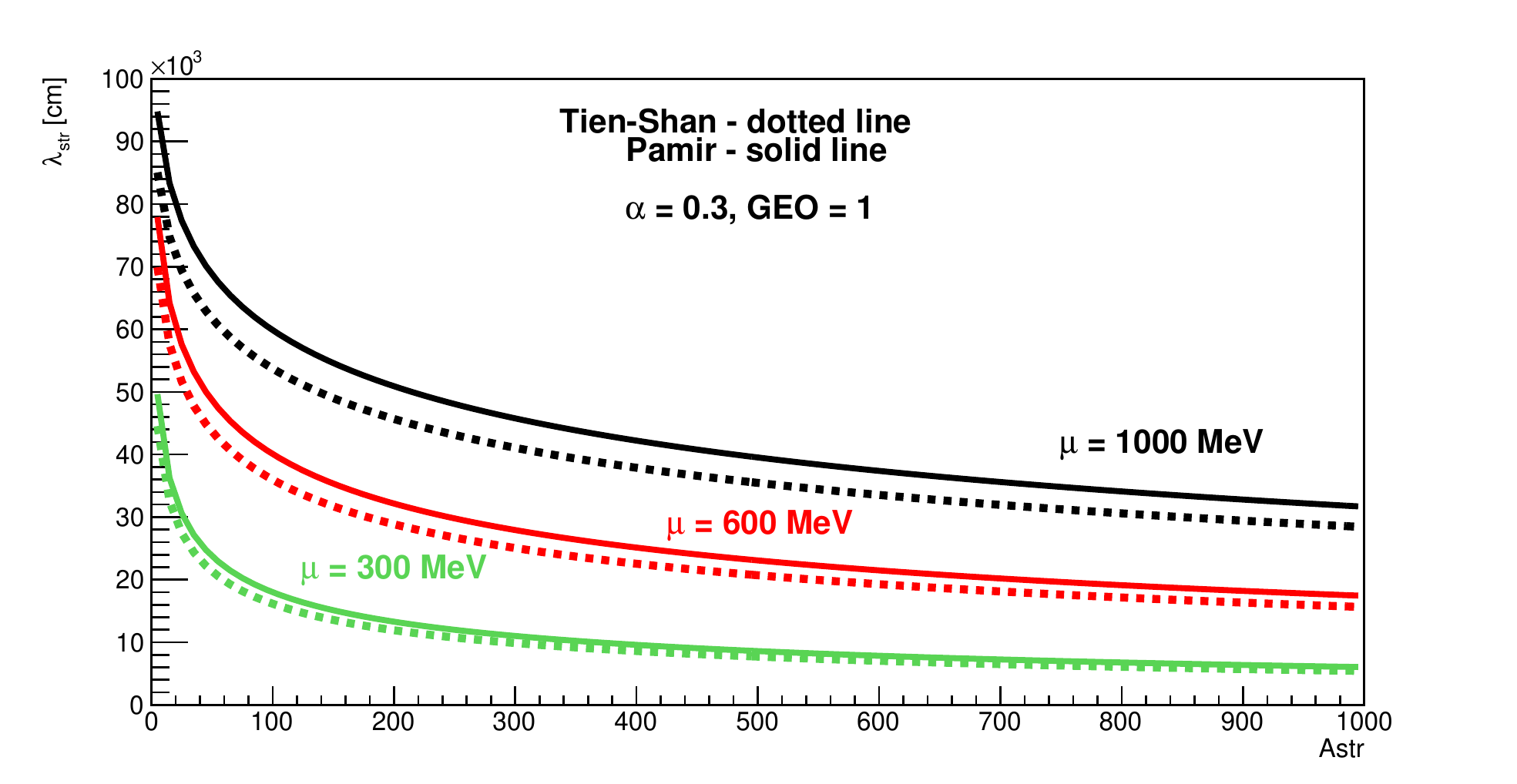}
\includegraphics[width=0.52\linewidth]{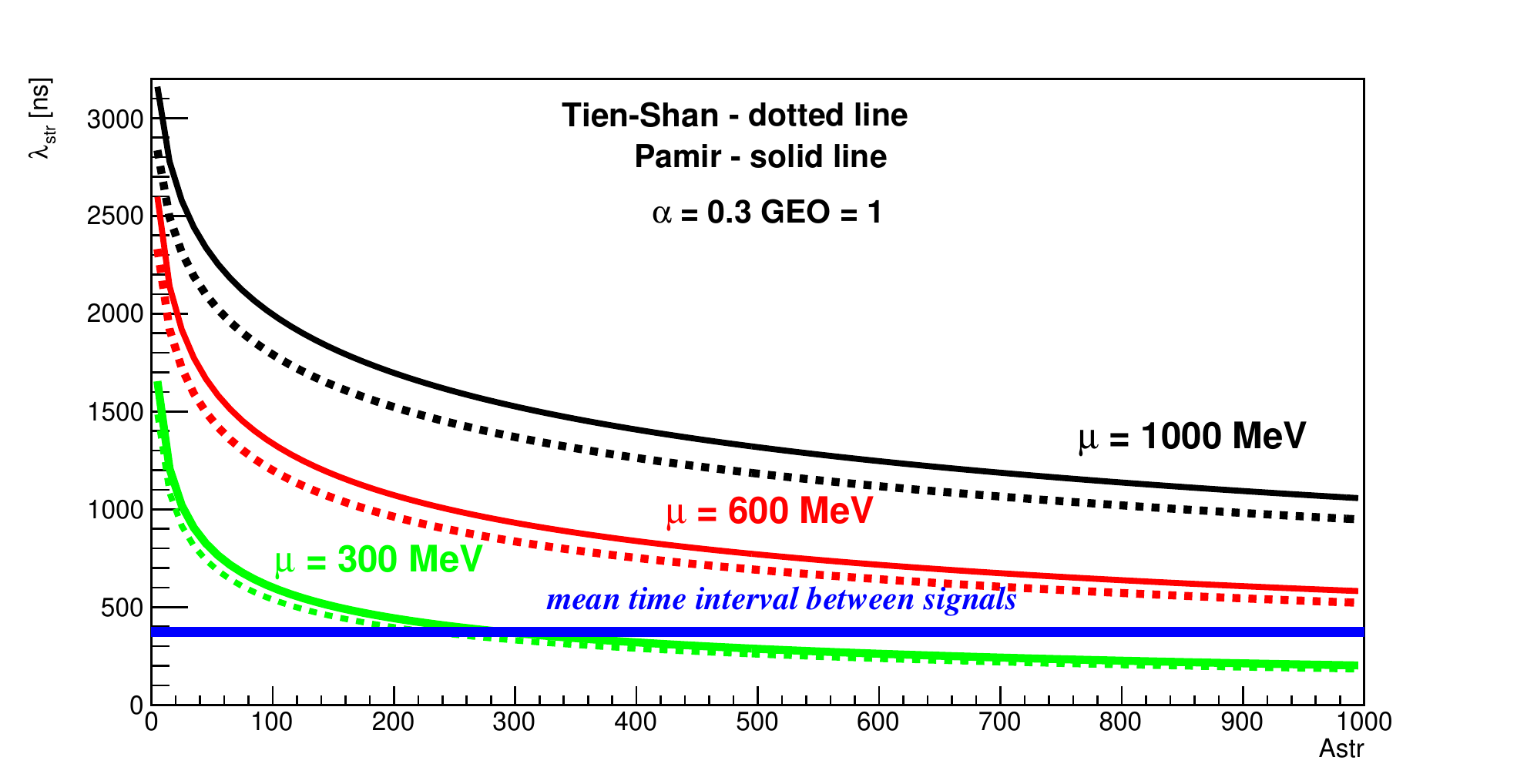}
\hspace*{-1cm}
\includegraphics[width=0.52\linewidth]{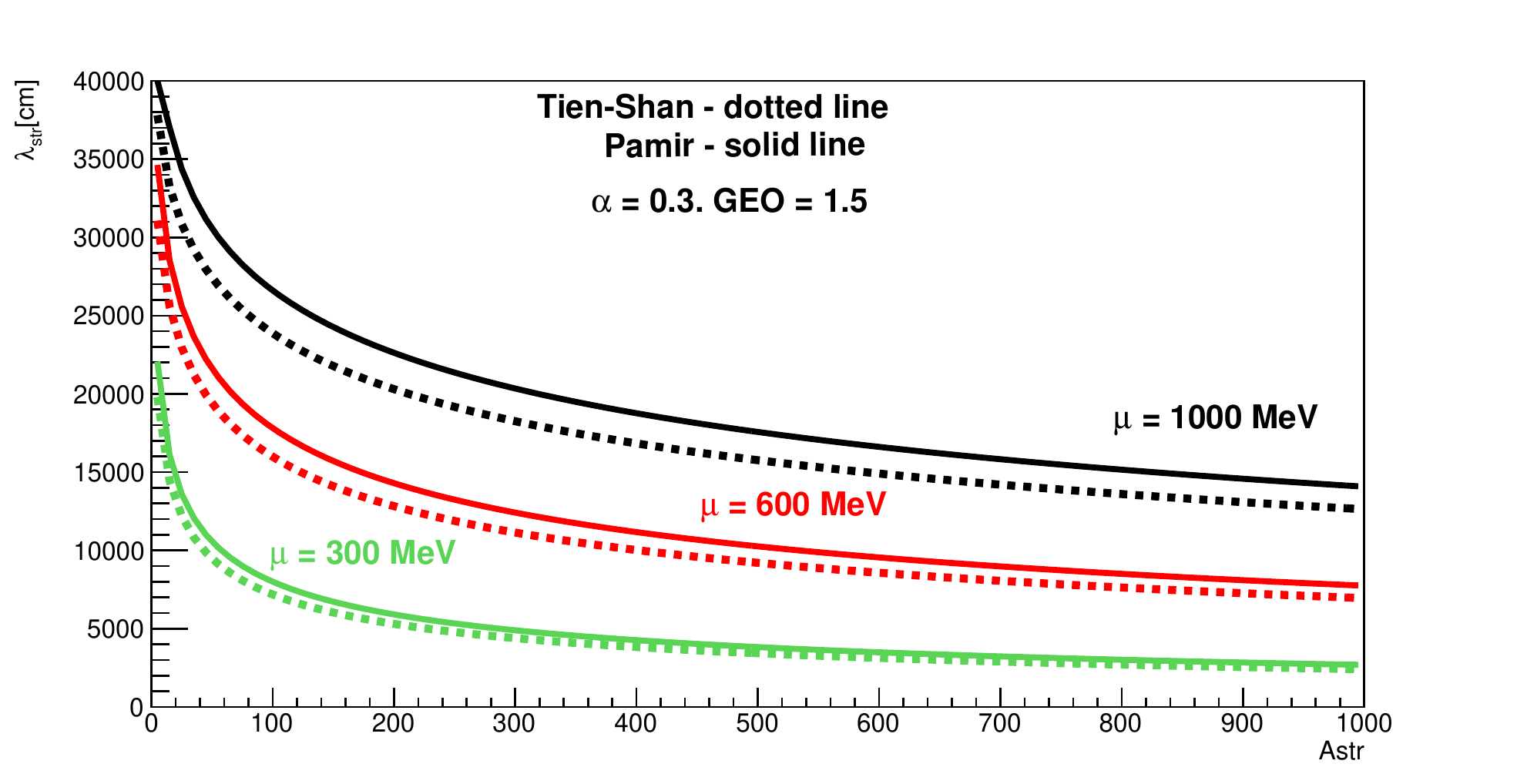}
\includegraphics[width=0.52\linewidth]{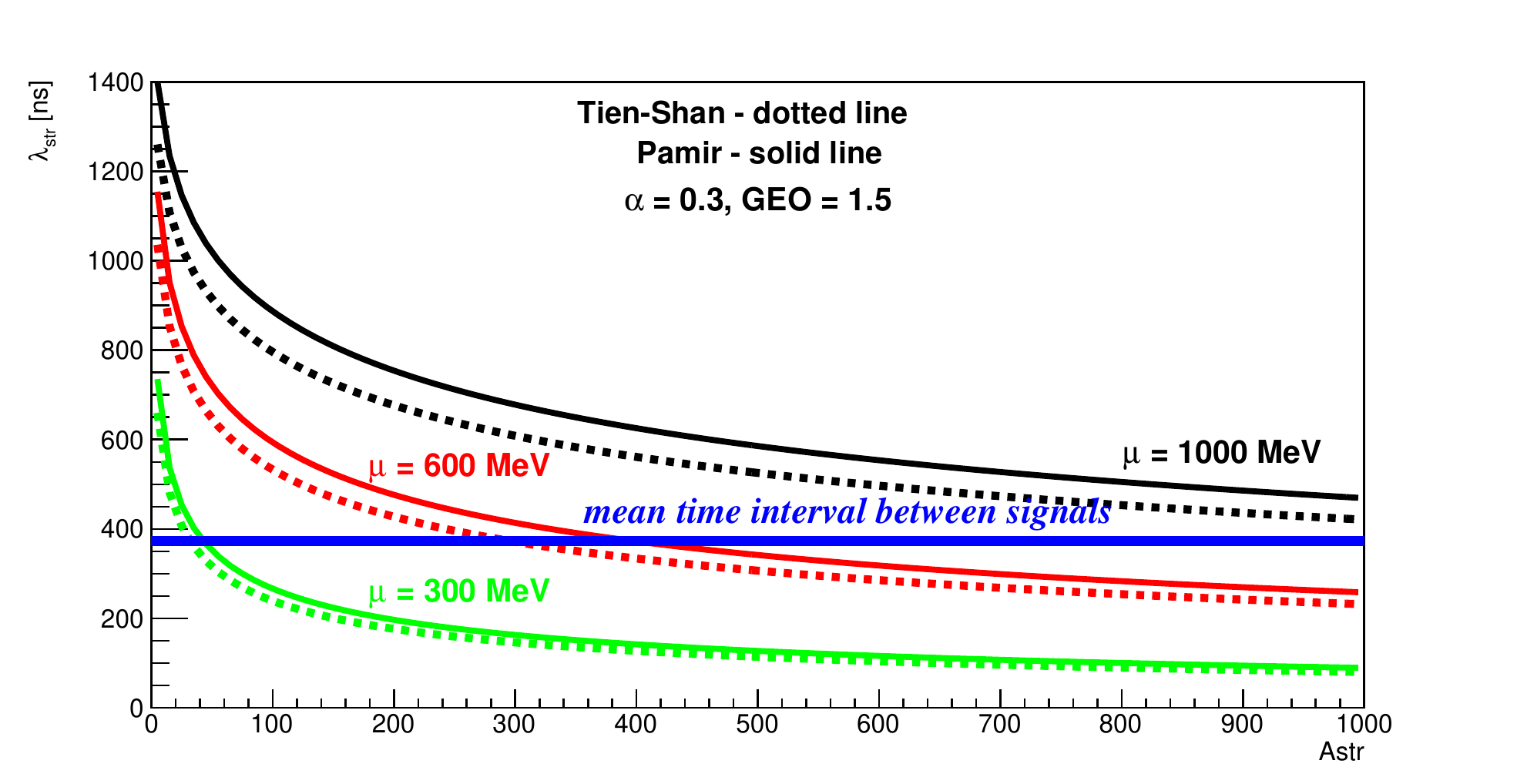}
\caption{Calculated $\lambda_{str}$ in the air as a function of a strangelet mass
 number $A_{str}$,  expressed in centimeters (left pictures) and 
in nanoseconds (right pictures). $\alpha_{s}$ = 0.3 and  GEO = 1 (upper pictures) or GEO = 1.5 (lower
pictures) were assumed. 
Comparison of $\lambda_{str}$ calculated for  
Pamir (solid line) 
and for Tien - Shan (dotted line)   is shown. The mean time distance $<\Delta\tau>$ between the peaks 
observed in the 5-peak MME event \cite{Beisembaev_5peaks} is indicated by the blue straight line.}
\label{fig:lambda_Pamir_TS_alfa03}
\end{center}
\end{figure}

There are, however, essential differences between both kind of experiments. 
The main difference between the Pamir and the Horizon experiments is that they detect nuclear electromagnetic
cascades developing in different absorbers. In the Pamir chambers we detect cascades developing in
Pb, being very heavy absorber. The signals detected  by the Horizon experiment are produced by  the 
nuclear and electromagnetic cascades developing in the air. The air is not only much lighter absorber
than Pb but additionally its density changes with the altitude. 
Fig.~\ref{fig:lambda_Pamir_TS_alfa03}
  shows  $\lambda_{str}$ in the air  as a function of a strangelet mass 
number $Astr$, calculated for $\alpha_{s}$ = 0.3, GEO = 1 and 1.5,  three values of $\mu$ = 300, 600 and 
1000 MeV, for Pamir and Tien-Shan altitudes. 
 Interaction lengths  of strangelets $\lambda_{str}$ have been
expressed in centimeters (left pictures) and in nanoseconds (right pictures), by assuming that
strangelets are flying with the time velocity $c$. The mean distance between the peaks 
observed in the 5-peak MME event \cite{Beisembaev_5peaks} is indicated by the blue straight line. 

\begin{figure}[t]
\vspace*{-0.5cm}
\begin{center}
\hspace*{-1cm}
\includegraphics[width=0.52\linewidth]{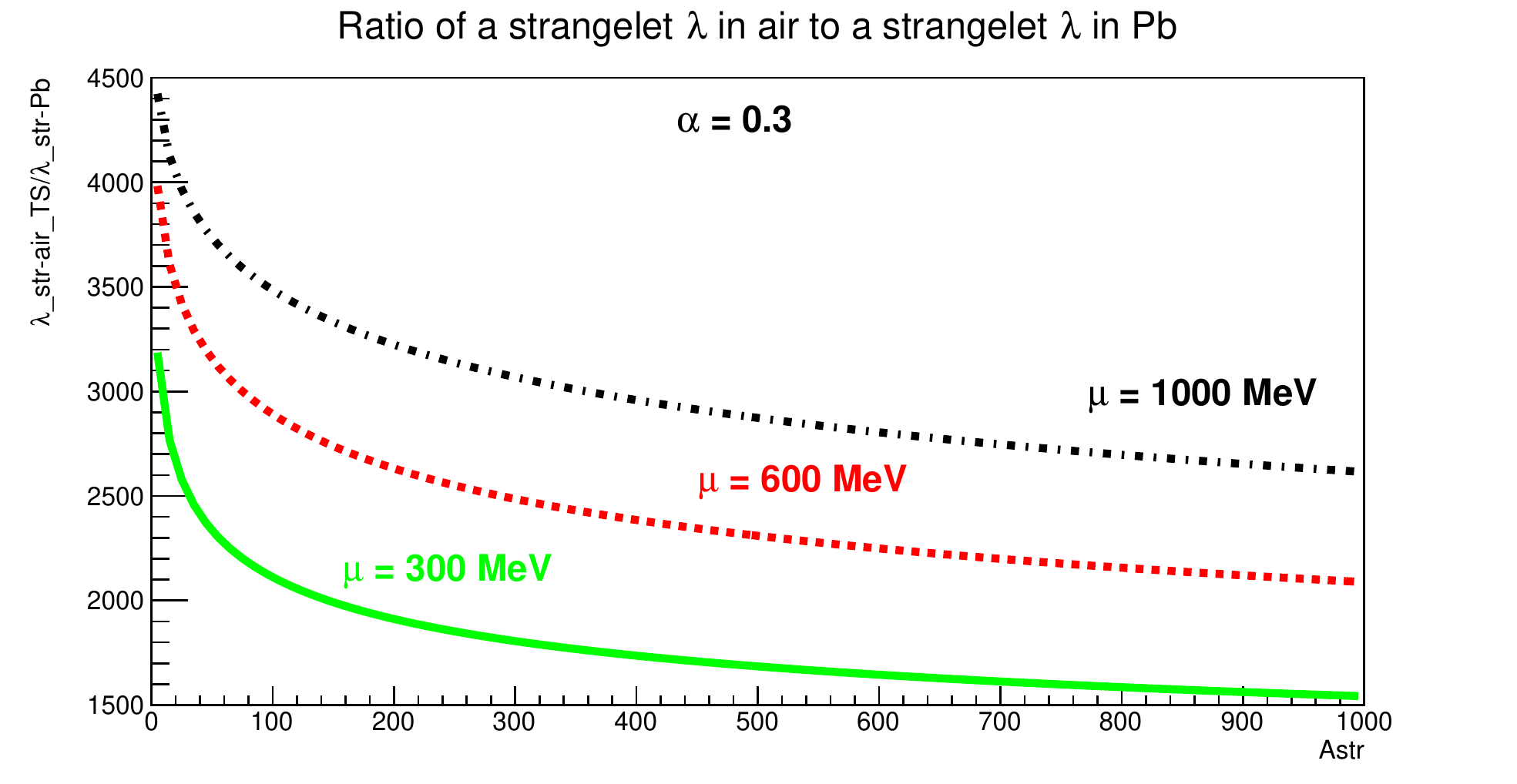}
\includegraphics[width=0.52\linewidth]{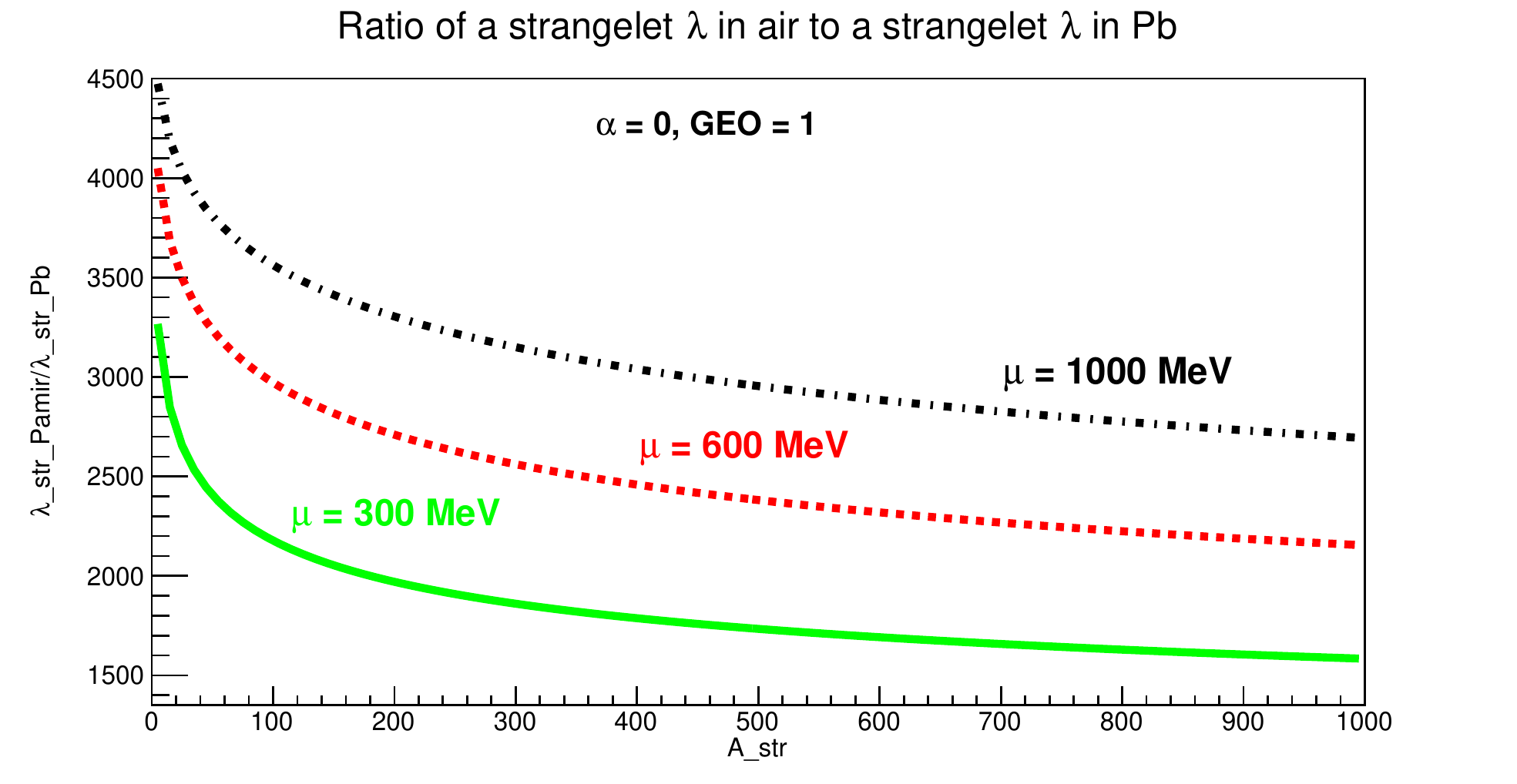}
\caption{Calculated ratios of  $\lambda_{str}$ [cm] in the air to these in the 
Pb absorber,
 at  Tien-Shan (left picture) and Pamir (right picture). GEO = 1 has been assumed in both cases.}
\label{fig:lambda_ratios}
\end{center}
\vspace*{-0.4cm}
\end{figure}

Fig.~\ref{fig:lambda_ratios} shows the ratios of calculated  $\lambda_{str}$
 for strangelets passing through
the air to these for strangelets passing through the Pb absorber.     
 
Our calculations indicate that: 

\begin{itemize}
\item The general behaviour of $\lambda_{str}$ as a function of $Astr$ is similar in both cases. 
 As well as for strangelets passing through the air and through the Pb  absorbers,
 $\lambda_{str}$ decreases with 
an increase of $Astr$ and
increases with $\mu$.

\item The calculated   $\lambda_{str}$ are longer for GEO = 1 than for GEO = 1.5. 
Dependence of $\lambda_{str}$
on $\alpha_{s}$ is very weak.
 
\item The mean interaction paths of strangelets  in the air are much longer than these  in Pb, i.e.
 $\lambda_{str-air} >> \lambda_{str-Pb}$.
For $0 < A_{str} < 1000$ their ratios are in the range $\sim$ 1500-4500.  
As well as the ratios of 
cross sections $\sigma_{str-air}/\sigma_{str-Pb}$ and the ratios of  
  $\lambda_{str-air}/\lambda_{str-Pb}$  do not depend on the geometrical factor GEO. 
They also depend very weakly on the $\alpha_{s}$ value.
         
\item The curves obtained for the Pamir and for the Tien-Shan altitudes  are close  one to the other.

 \item 
As it is seen from the position  of the blue straight lines 
(Fig.~\ref{fig:lambda_Pamir_TS_alfa03})
the experimentally found time distances $\Delta\tau$  between arrivals of
the  successive signals are within the calculated values of  $\lambda_{str}$ (with exception
of very low mass strangelets).  
 It indicates that {\em the calculated   $\lambda_{str-air}$  are related 
 to the observed values of time distances between the successive pulses of many-maxima signals
detected by the HORIZON experiment.
It is the same conclusion as that drawn from our earlier analysis of exotic many-maxima cascades
registered in the thick Pb chambers.}  
\end{itemize}
  
\subsection{Comparison of the Pamir Pb chambers and the Horizon Tien Shan detectors}

As it has been already mentioned in the previous subsection the main difference between 
both detectors is the kind of absorbers. In the Pamir Pb chambers the mean interaction path of a strangelet
$\lambda_{str-Pb}$ 
is much shorter than  $\lambda_{str-air}$ in the air being the absorber for the Horizon detectors.

 In a consequence
we are able to observe directly each  point of a  strangelet interaction with the Pb nucleus and 
directly watch the strangelet passage through the Pb chamber. The nuclear electromagnetic cascades,
 developing in Pb,
after each strangelet interaction are visible at X-ray film detection layers in the 
form of black spots, of the radius of only several hundreds of microns.  
The distances between the starting points of the successive cascades are directly
connected  with the length of  strangelet interaction paths $\lambda_{str-Pb}$. The 60 cm Pb thickness of the 
chamber corresponds to
$\sim  3.5 \lambda_{int}$ for nucleons. In the case of strangelets, characterized by the shorter 
 interaction paths, the chamber is a suitable tool for  detection  and 
 for watching of  many consecutive strangelet interactions.
   
In the case of strangelets passing through the air layer the situation is different. 
The long $\lambda_{str-air}$ causes that the air cascades developing  
between the consecutive  strangelet interactions create the EAS with large as well as 
longitudinal and lateral sizes.
The direct measurements of distances between the successive strangelet  interactions are 
impossible because the  developing
EAS is detected at the only one  level - at the Tien-Shan detection station altitude.   
However,  time distances $\Delta \tau$ between arrivals of  signals from the subsequent  
interactions are connected with the 
 mean interaction paths of the objects responsible for their production. 
The knowledge of $\lambda_{str-air}$ 
  enables to estimate  a time of strangelet  interactions with the air nuclei.

As it is seen from Fig.~\ref{fig:lambda_Pamir_TS_alfa03}  
 the experimental  $ <\Delta \tau>$ value  is located between the calculated curves,
 what  indicates  a consistency of assumed by us scenario with the
experimental observations. However, to draw the more detailed conclusions, 
  a dependence of $\Delta \tau$ 
on a  distance  $R$ should be studied.  This question will be discussed
in the next section.          

At the moment the next essential point should be mentioned. The considered detectors are located at two 
different
altitudes.
The difference of their altitudes $\Delta H_{Pamir-TS} \approx 100 g/cm^{2}$. 
 Expressing $\Delta H$ in a number
of  strangelet mean interaction paths $N[\lambda_{str-air}]$ we found that for $\mu \approx$ 300 - 600  MeV and 
for  
strangelets with $Astr \geq 100$ this number is close  
to the thickness of 60 cm  Pb chamber $L_{Pb}$,
 i.e. {\bf $\Delta H_{Pamir-TS}[\lambda_{str-air}] \approx L_{Pb}[\lambda_{str-Pb}]$}.
So, {\bf we can expect the similar mean number of the consecutive strangelet interactions  in
both detectors.}    
For example, for strangelets with $Astr$ = 100 (500), $\mu$ = 300 MeV and  $\alpha_{s}$ = 0.3
 the thickness of the Pamir 
Pb chamber $L_{Pb} \approx$ 8 (13) $\lambda_{str-Pb}$ and the thickness of the air layer between Pamir and Tien-Shan
stations $\Delta H_{Pamir-TS} \approx$ 6 (13)  $\lambda_{str-air}$. 

\begin{figure}
\begin{center}
\includegraphics[width=0.49\linewidth]{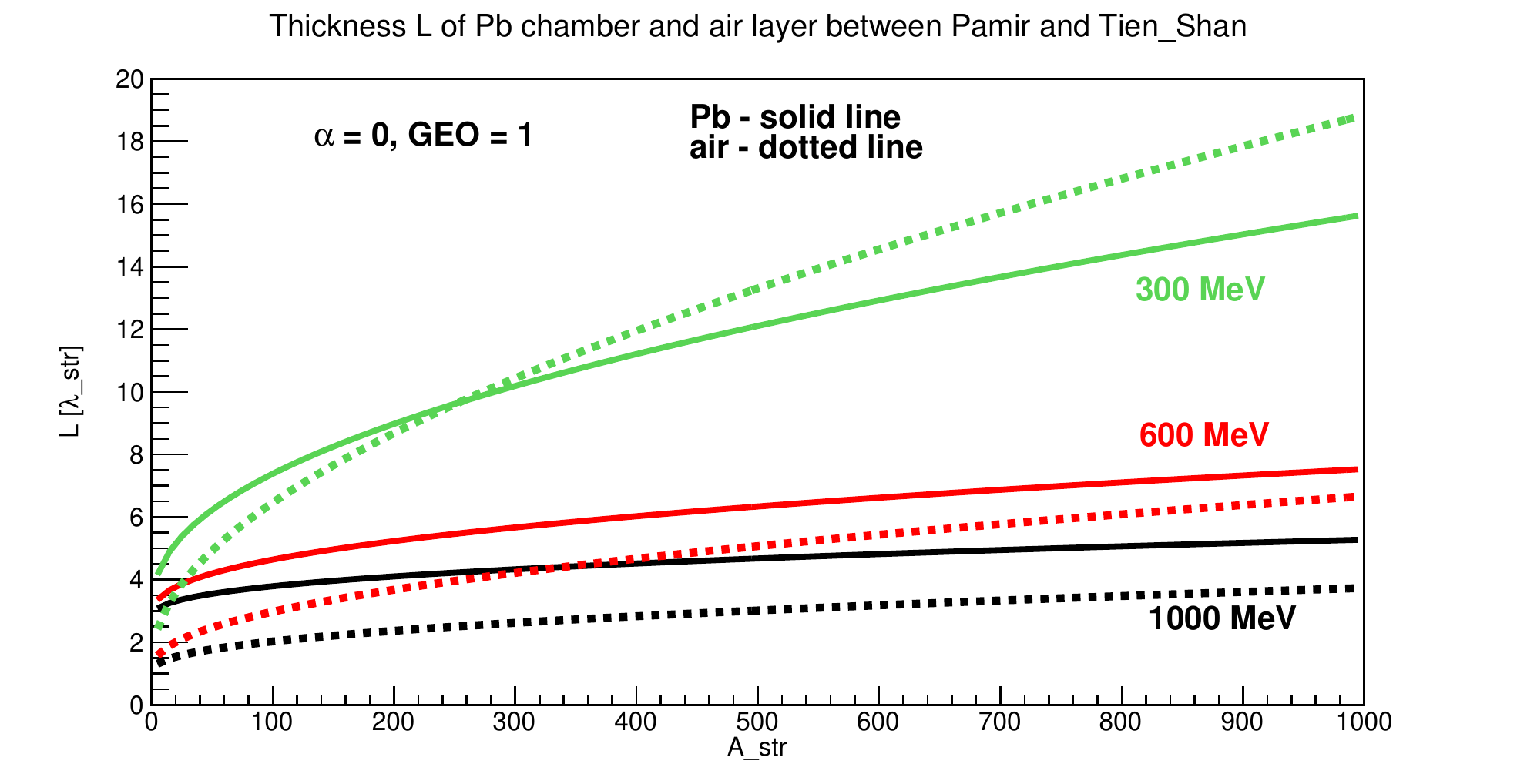}
\includegraphics[width=0.49\linewidth]{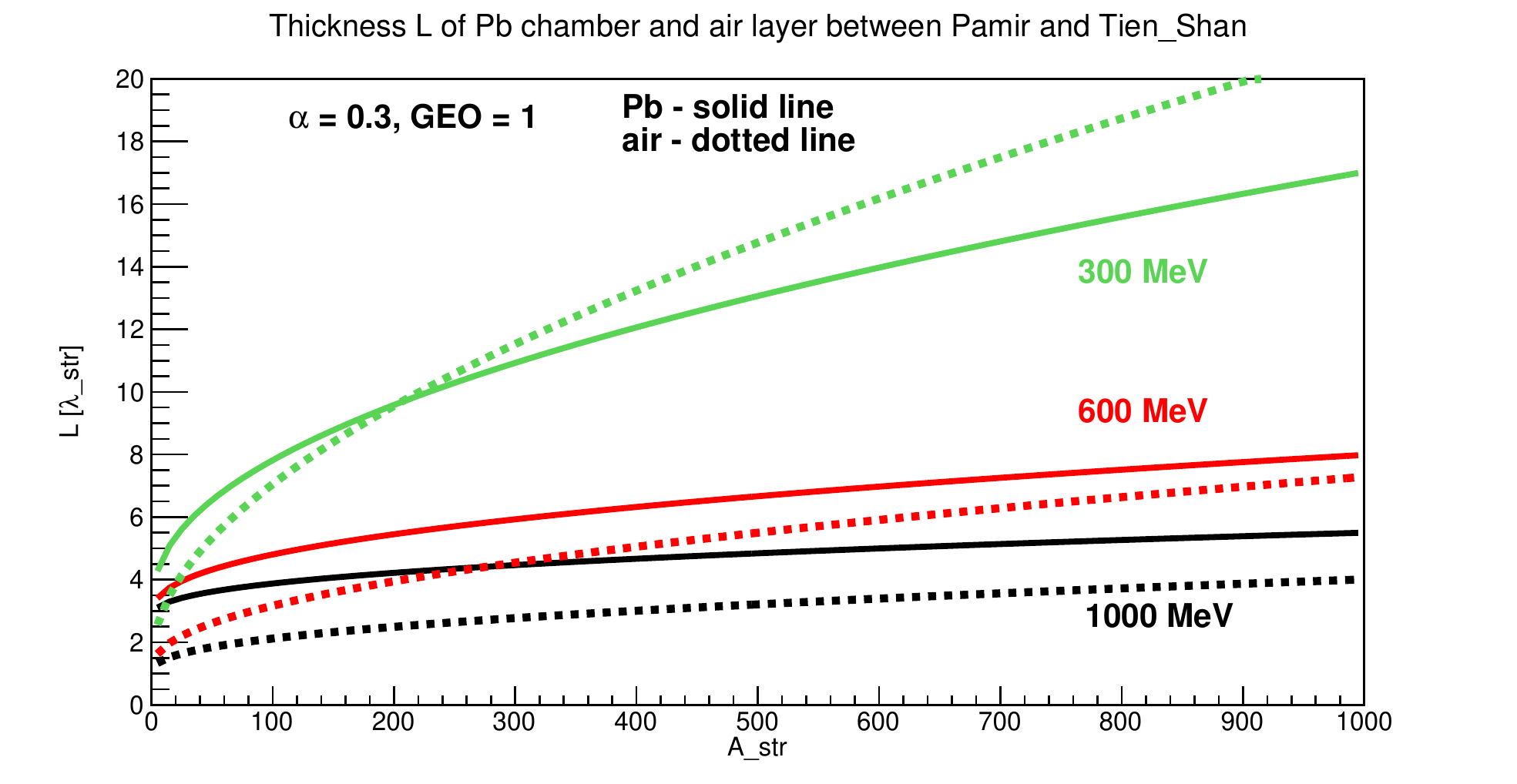}
\includegraphics[width=0.49\linewidth]{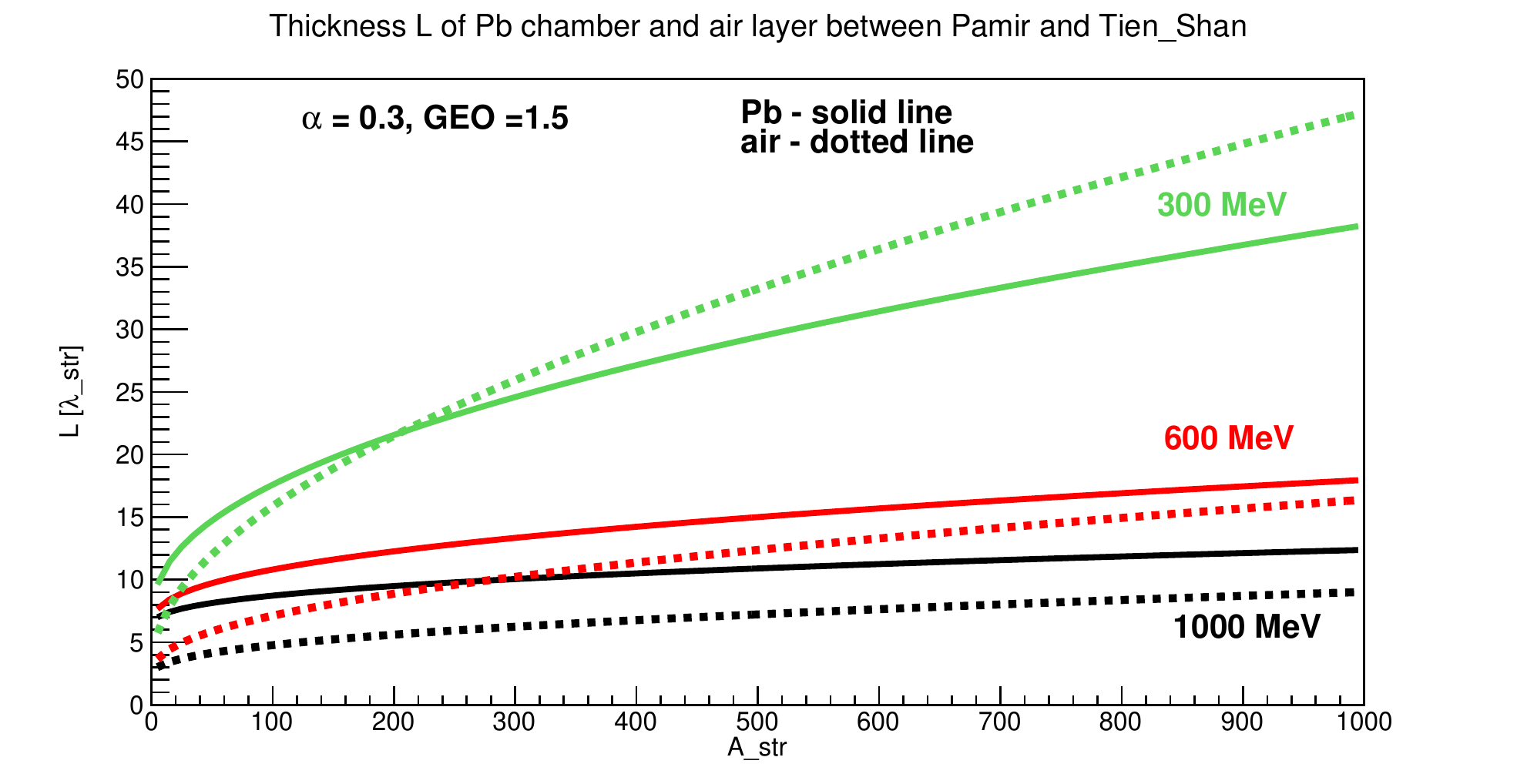}
\includegraphics[width=0.49\linewidth]{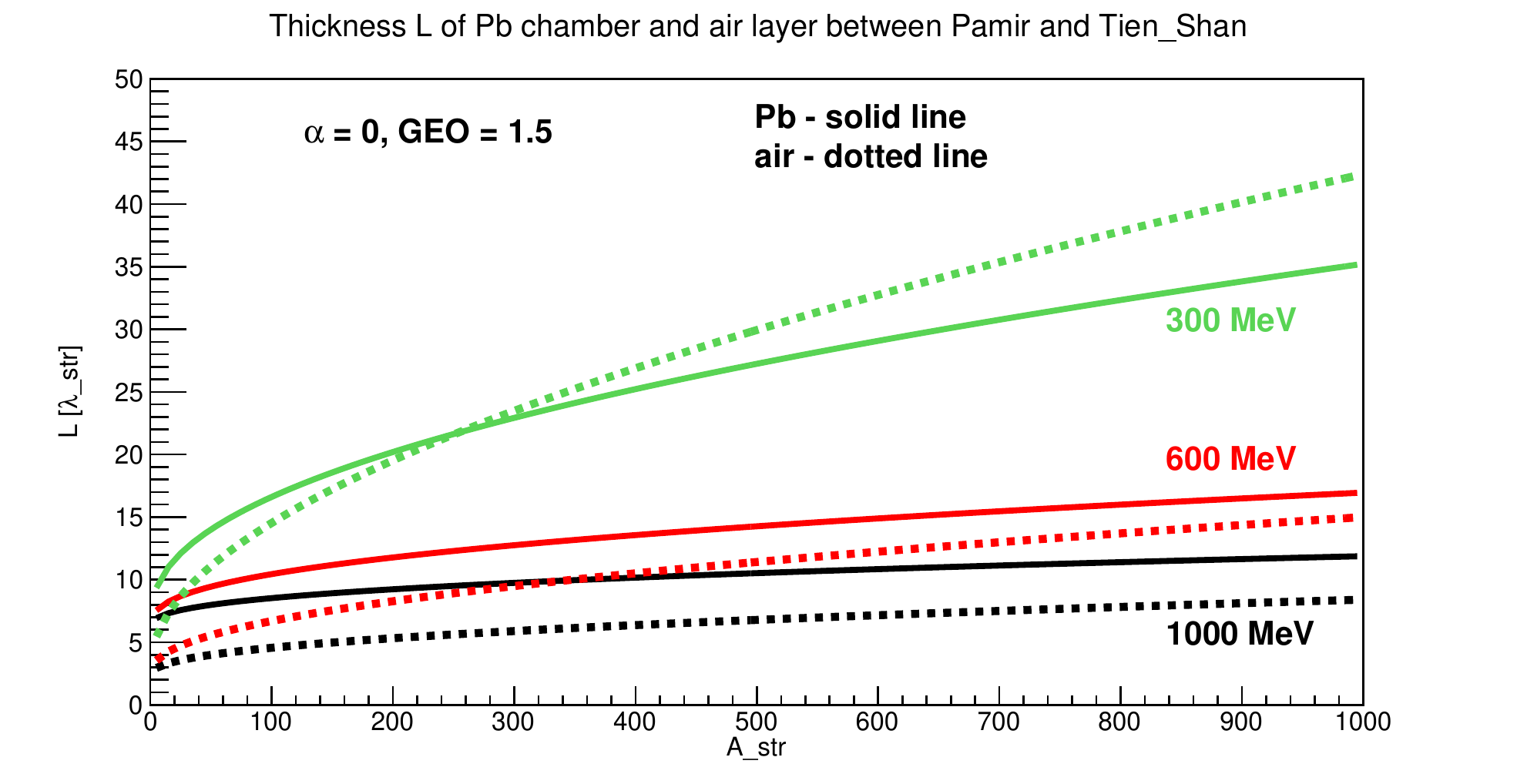}
\caption{Thicknesses L [$\lambda_{str}$] of the  60 cm Pb chamber and 100 g/cm$^{2}$ of the air layer between
the Pamir and Tien-Shan stations, expressed in a number of  strangelet mean interaction paths  $\lambda_{str}$.}
\label{fig:Thicknesses_100gcm2air}
\end{center}
\vspace*{-0.4cm}
\end{figure}

\begin{figure}
\vspace*{-2cm}
\begin{center}
\includegraphics[width=0.49\linewidth]{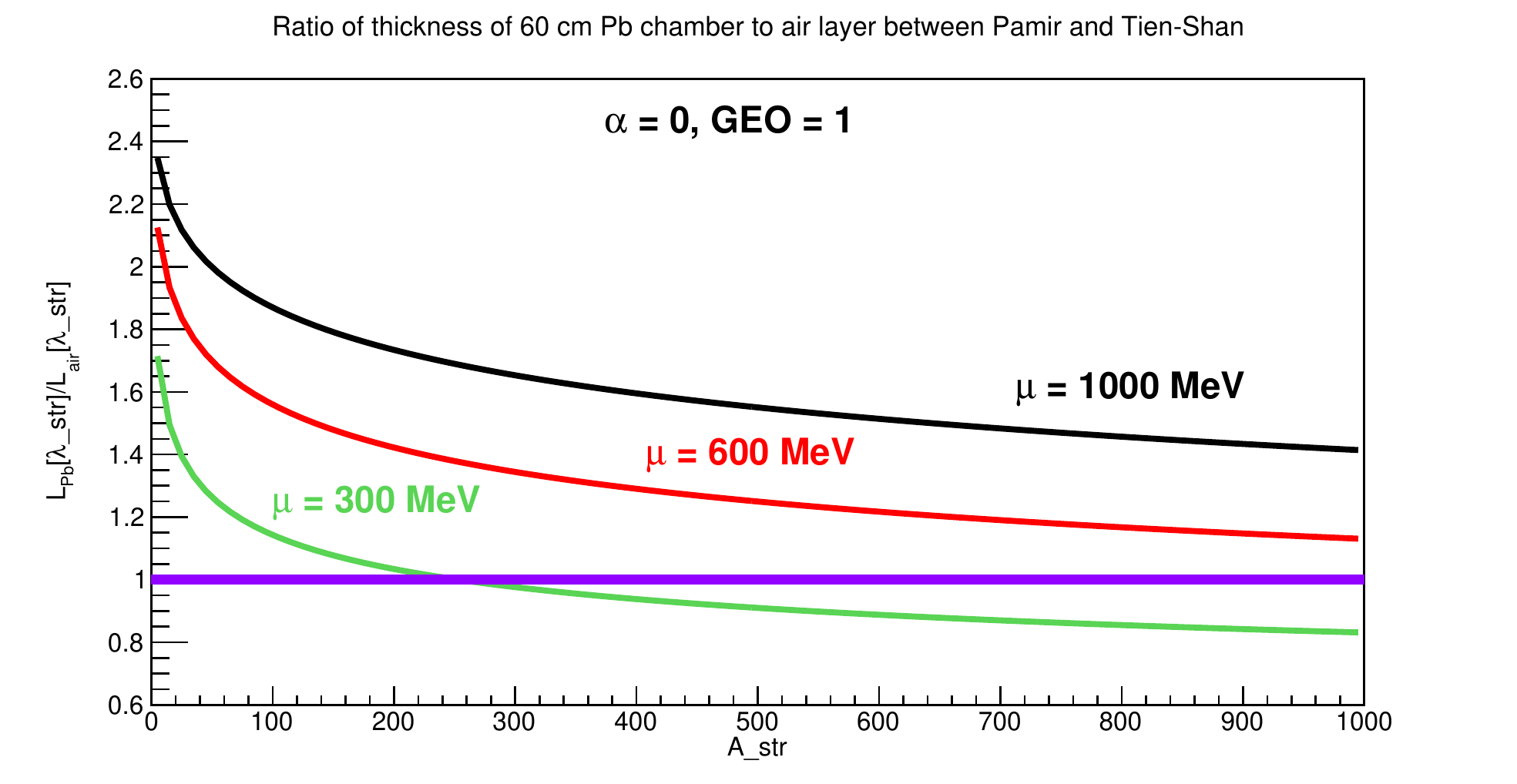}
\includegraphics[width=0.49\linewidth]{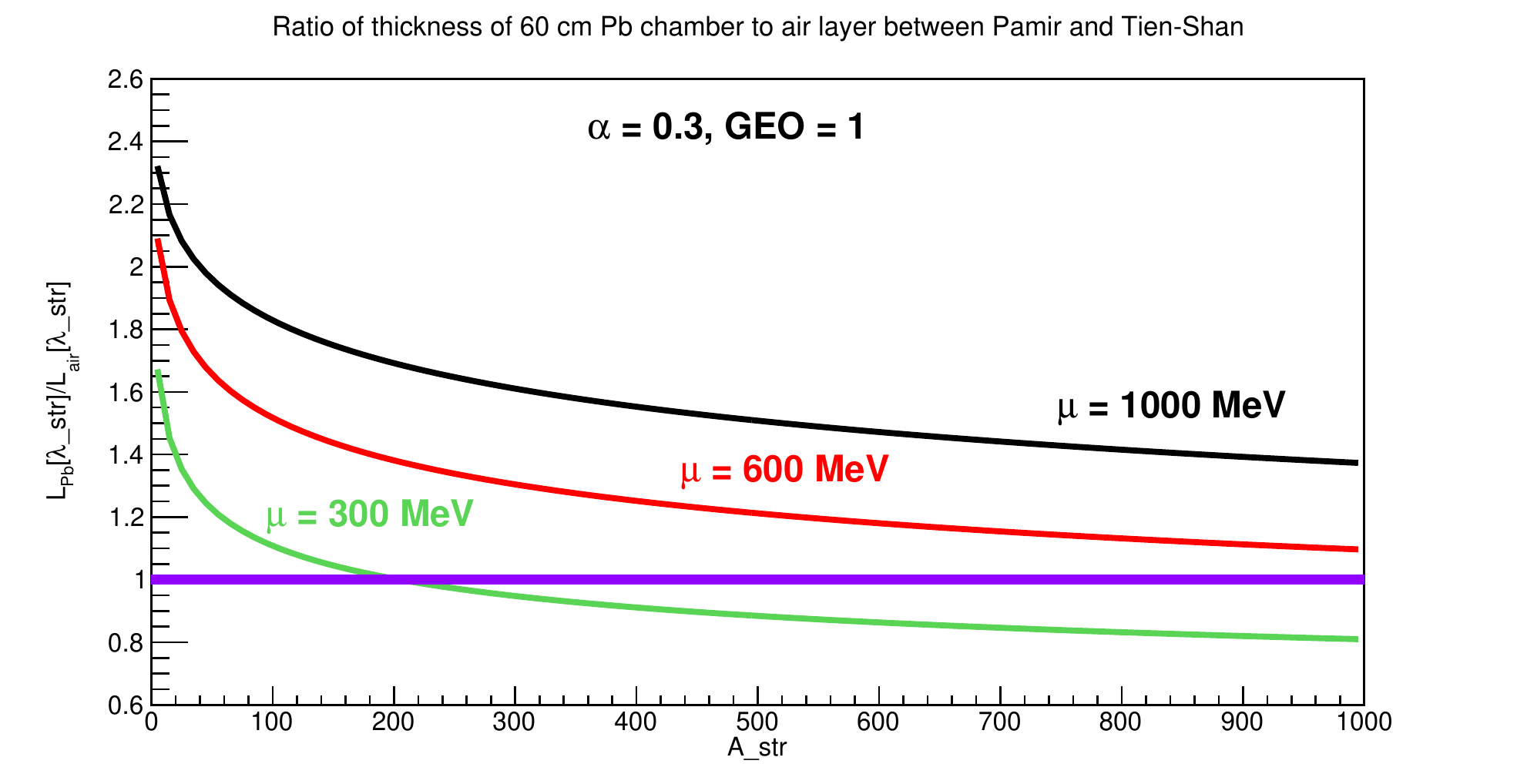}
\includegraphics[width=0.49\linewidth]{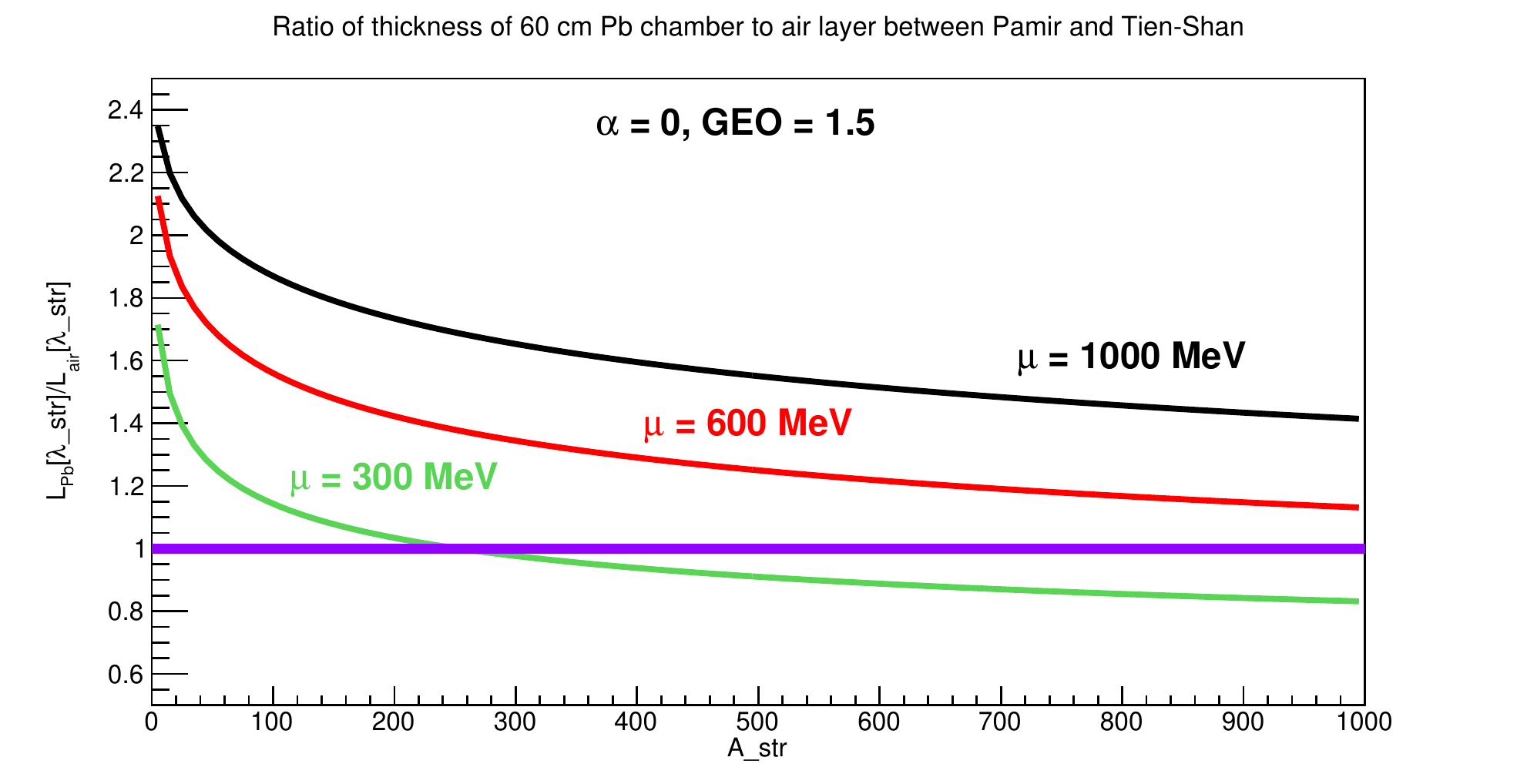}
\includegraphics[width=0.49\linewidth]{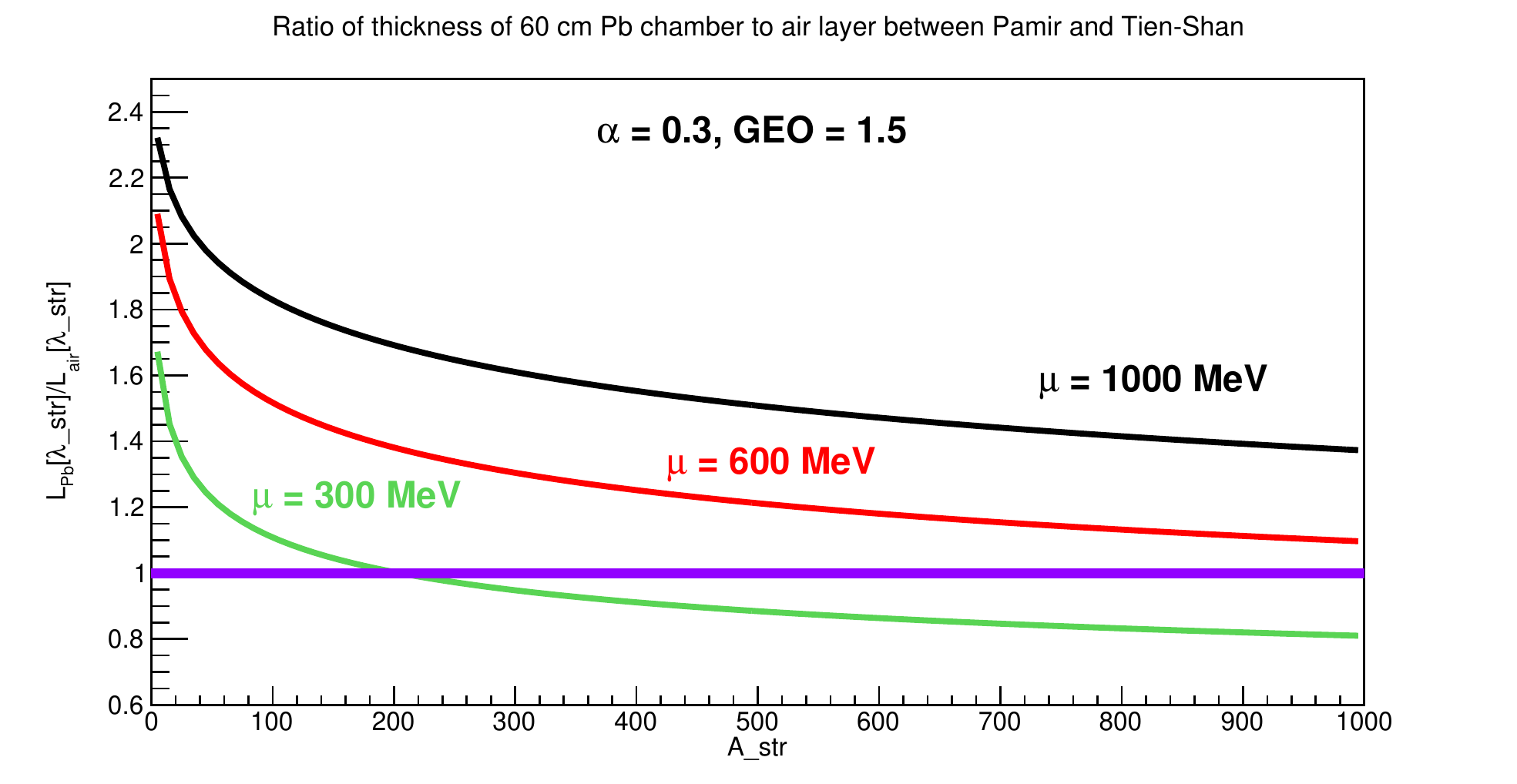}
\caption{Ratio of thicknesses: 60 cm Pb chamber to 100 g/cm$^{2}$ of the air layer between
the Pamir and Tien-Shan stations, expressed in  $\lambda_{str}$.}
\label{fig:Ratios}
\end{center}
\vspace*{-1cm}
\end{figure}

For details see 
Figs.~\ref{fig:Thicknesses_100gcm2air} and  ~\ref{fig:Ratios}.

Fig.~\ref{fig:Thicknesses_100gcm2air} shows thicknesses of the  
60 cm Pb chamber $L_{Pb}[\lambda_{str}]$ (solid lines) 
 and the 100 g/cm$^{2}$  air layer between
the Pamir and Tien-Shan stations $L_{air}$[$\lambda_{str}$] (dotted lines),
 as a function of $Astr$.  
The curves calculated for 3 values of a strangelet quark-chemical potentials are marked by different colours.
The solid and the dotted curves of the same colour are close one to the other what  proves
 a similar thickness
of both detectors. 

The ratios of the both absorber thicknesses are shown in Fig.~\ref{fig:Ratios}. The positions of
the blue straight lines, drawn at $L_{Pb}[\lambda_{str}]/L_{air}[\lambda_{str}] = 1$, show 
that for reasonable values of 
strangelet parameters ($Astr > 100$, $\mu \sim 300-600$ MeV)
 the thickness of the  Pb chamber is close or only a little greater than the thickness of the air layer
between the Pamir and Tien-Shan stations.

For illustration see the sketch drawn in Fig.9. 

So, we can expect that:
\begin{itemize}
\item  Almost the same kind and the same number  of cosmic ray species which is falling
and interacting in the Pamir chambers is reaching  the Horizon detectors
\item  In the case of 
the strangelet projectiles with the same characteristics we can expect 
almost {\em the same mean number of the consecutive strangelet interactions in both detectors.}
\end{itemize}

\begin{figure}[t]
\vspace*{-0.5cm}
\begin{minipage}{0.6\linewidth}
\includegraphics[scale = 0.35]{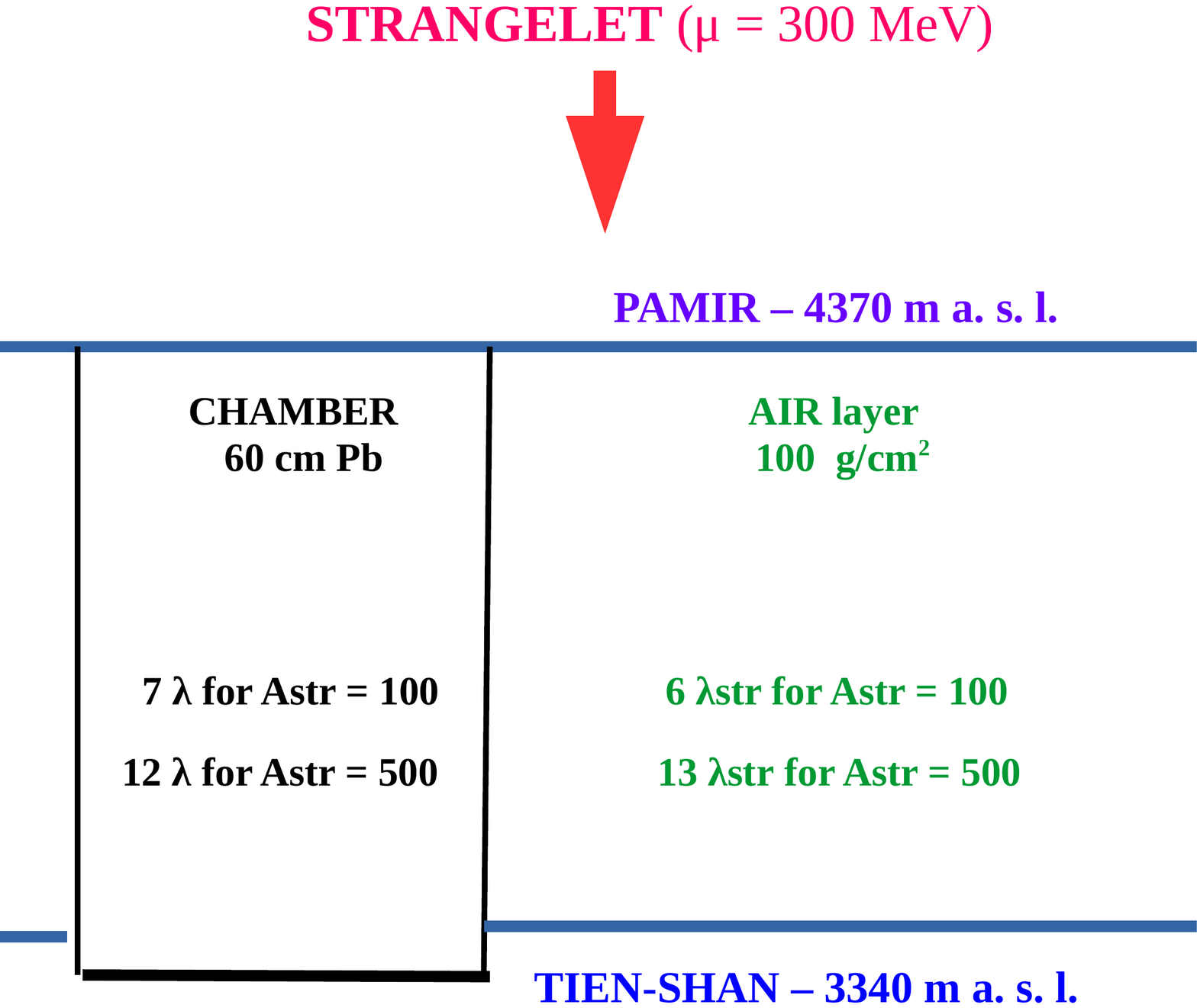}
\end{minipage}
\hspace*{-1cm}
\vspace*{-2cm}
\begin{minipage} {0.38 \linewidth}
\caption{Illustration and comparison of Pamir and Horizon
detector  abilities to a cosmic ray strangelet detection}
\end{minipage}   
\label{fig:szkic-thickness-air-Pb}
\vspace*{-1cm}
\end{figure}

 It should be, however,  mentioned that the Pb chambers are more  suitable for observation and investigation 
of
low mass strangelets in contrary to  the Horizon detectors which seem to be  more appropriate  for 
detection of  high mass strangelets.

\subsection{Proposed explanation and its correspondence to the experimental data}

\subsubsection{General idea}

For explanation of the multimodal events we propose the simplified picture of "stable" strangelet interactions
 in the air. The mechanism is the same as that we used  and described in \cite{strangelet_Pb} 
for explanation of the  strongly penetrating 
many maxima cascades detected in the Pamir Experiment.
Fig. 10 illustrates the idea.

\begin{figure}
\vspace*{-3cm}
\begin{minipage}{0.6\linewidth}
\includegraphics[scale = 0.35]{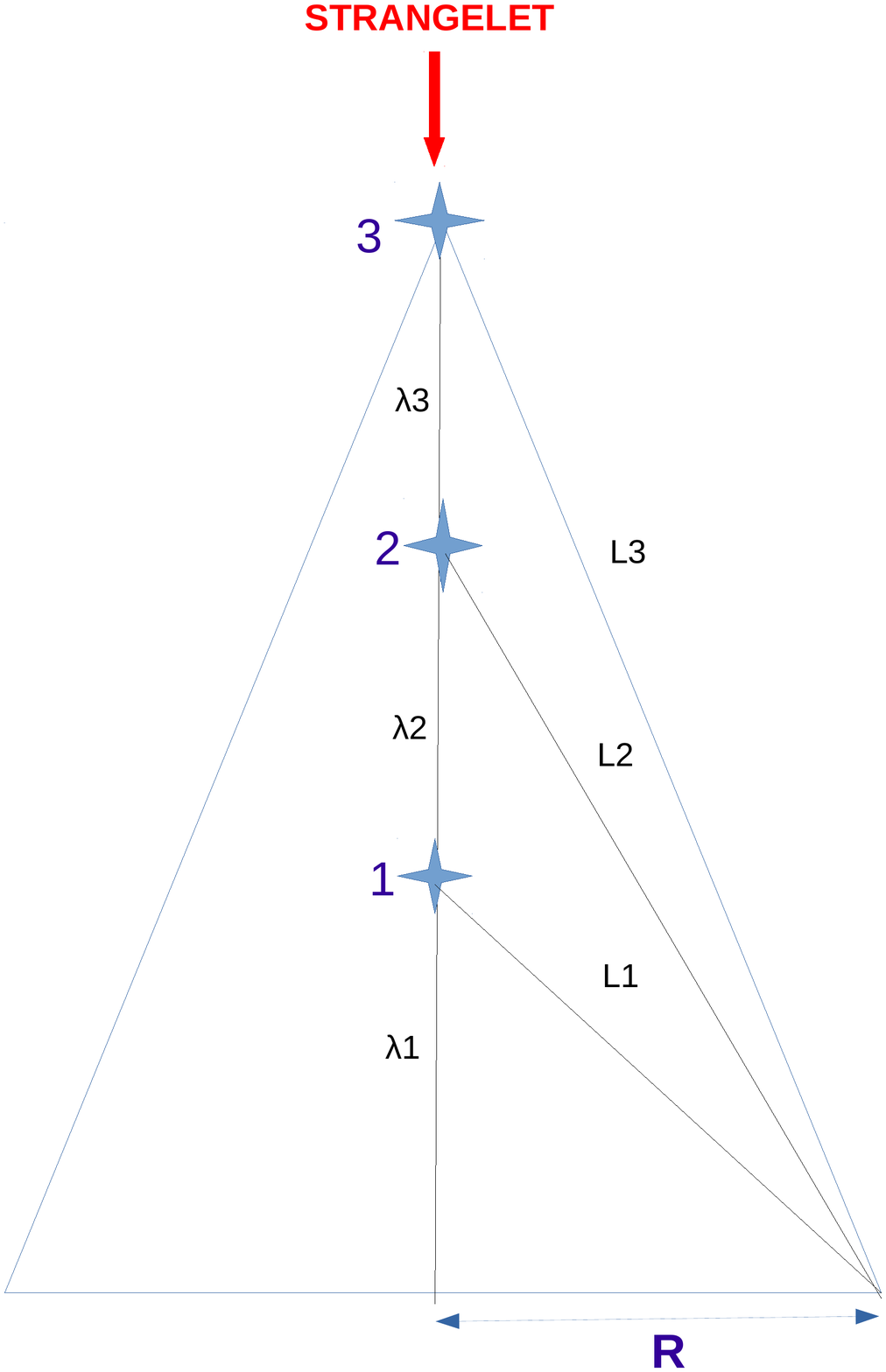}
\end{minipage}
\vspace*{-2.5cm}
\hspace*{-1.5cm}
\begin{minipage}{0.3\linewidth}
\caption{Scheme  of successive strangelet interactions in the air responsible for
  the appearance of many peak structures in detectors. Only 3 interactions are drawn for illustration.}
\end{minipage}
\label{fig:szkic_idea}
\vspace*{2.cm}
\end{figure}

In our simplified picture we assume that:
\begin{itemize}
\item  Each of $n$ signals seen by the Horizon  detector located at the  same  distance $R$ from the EAS axis 
comes from the successive strangelet
interaction with the air nuclei. The most delayed peak, marked here by no 1 comes from a strangelet interaction at the altitude 
  $\lambda_{1}$ above the detection level, peak no 2 from a strangelet interaction at the altitude
  $\lambda_{1} + \lambda_{2}$, peak no 3 from a strangelet interaction at the altitude  $\lambda_{1} + \lambda_{2} + \lambda_{3}$
 etc.
\item The distances between the successive strangelet interactions
 are equal to the values of the corresponding strangelet interaction paths.

 $\lambda_{1} \approx \lambda_{2} \approx \lambda_{3}...\approx \lambda_{n} \equiv \lambda \equiv \lambda_{str}$,
where $\lambda$ is the mean value of the strangelet interaction paths for the detected collisions.
\item  As well as the strangelet and the resulting nuclear-electromagnetic cascades fly with the light 
velocity $c$.
\end{itemize}
 Expressing $\lambda_{str}$ in the time units   we can calculate
 the time distances $\Delta\tau_{n(n-1)}$ 
between arrival of signals from the successive strangelet  interactions. For example, if $n$ = 3 we have:

\begin{enumerate}
\item time distance  between interactions at the height $H_{2} = 2\lambda$ and $H_{1} = 1\lambda$:\\
 $\Delta\tau_{21} \approx \lambda +  \sqrt{(\lambda^{2} + R^{2})} - \sqrt{(2\lambda)^{2} + R^{2}}$ 

\item time distance between interactions at the height $H_{3} =  3\lambda_{3}$ and $H_{2} = 2\lambda$:\\
 $\Delta\tau_{32} \approx \lambda + \sqrt{(2\lambda)^{2}  + R^{2}} - \sqrt{(3\lambda)^{2} + R^{2}}$
\end{enumerate}     

We see that  $\Delta\tau$ depends as well as on the interaction height $H$, the values of
the strangelet interaction paths and on
the distance $R$ of the detector from the EAS axis. 
 
 The following limit values of $\Delta\tau$ can be expected:
\begin{enumerate}
\item  for $R = 0$ or $R <<  \lambda$:\\
$\Delta\tau_{21} \rightarrow 0$, $\Delta\tau_{32} \rightarrow 0$, $\Delta\tau_{43} \rightarrow 0$ ... $\Delta\tau_{n(n-1)} \rightarrow 0$

\item for  $R >> n \lambda $:\\
$\Delta\tau_{21} \rightarrow \lambda_{2}$, $\Delta\tau_{32} \rightarrow \lambda_{3}$, $\Delta\tau_{43} \rightarrow
 \lambda_{4}$ ... 
$\Delta\tau_{n(n-1)} \rightarrow \lambda_{n} \equiv \lambda$

\item  for $R \approx \lambda$:\\
$\Delta\tau_{21} \approx \lambda(1 + \sqrt{2} - \sqrt{5}) \simeq 0.178 \lambda$ \\ 
$\Delta\tau_{32} \approx \lambda(1 + \sqrt{5} - \sqrt{10}) \simeq 0.073\lambda$ \\
$\Delta\tau_{43} \approx \lambda(1 + \sqrt{10} - \sqrt{17}) \simeq 0.030\lambda$ \\ 
....\\
$\Delta\tau_{n(n-1)} \rightarrow 0$ when $n \rightarrow \infty$ \\
i.e. $\Delta\tau$ is decreasing  with the increase of the successive strangelet interaction altitude.  

\end{enumerate}   

In this picture  the first detected signal comes from the first strangelet 
interaction, i.e. the interaction at the highest altitude. The last signal comes from the 
strangelet interaction at the height $H \approx \lambda_{n}$ from the detection level. {\bf In accordance 
with the experimental observations,
time distances $\Delta\tau$ between the successive signals are decreasing with the increase 
of the strangelet interaction altitude} (i.e. $\Delta\tau$ are the longest for the most delay signals, coming from interactions 
close to the detector level). This effect is additionally enhanced by the increase of the $\lambda_{str}$ value 
 with the decrease 
of $Astr$ in each of the successive interaction. 

The experimental MME events, shown in the section 2.2 have been detected at rather long distances $R$ from the EAS axis. 
$R$ in the range of $\sim$ 300-1000 m from the EAS axis means that we are somewhere between situations
described in the points 2 and 3. For the 5-peak MME event (Fig.~\ref{fig:MME_5peaks}) 
 detected at the very long distance ($R \simeq 898$ m) from the EAS axis we can expect that the criterion 2
($R >> \lambda$) is fulfilled, so the observed $\Delta\tau$ are close to the $\lambda_{str}$ values (see the blue line at
Fig.~\ref{fig:lambda_Pamir_TS_alfa03}).
For explanation of the events detected closer to the EAS axis the situation described in the point 3 ($R \simeq \lambda$) 
is  more appropriate.

 The assumed picture allows to understand the strange behaviour of MME events.
The most important statement is that {\bf the multipeak structure of signals could be the result of
the successive strangelet interactions in the air. The observed increase of the time intervals $\Delta\tau_{n(n-1)}$ between 
two successive signals is caused by two factors: the above described geometry  and additionally by a dependence of $\lambda_{str}$ on 
$A_{str}$.} As it has been shown in the section 4.2, $\lambda_{str}$ is longer for the low mass strangelets.
 As the  mass number of a strangelet $A_{str}$
 is decreasing in each act of interaction,  $\lambda_{str}$ is increasing with the number of collisions. 
 It directly
results in elongation 
of the measured time distances between two successive signals $\Delta\tau_{n(n-1)}$ with
the time of their arrival.

\begin{figure}
\vspace*{-1.5cm}
\begin{center}
\includegraphics[width=0.48\linewidth]{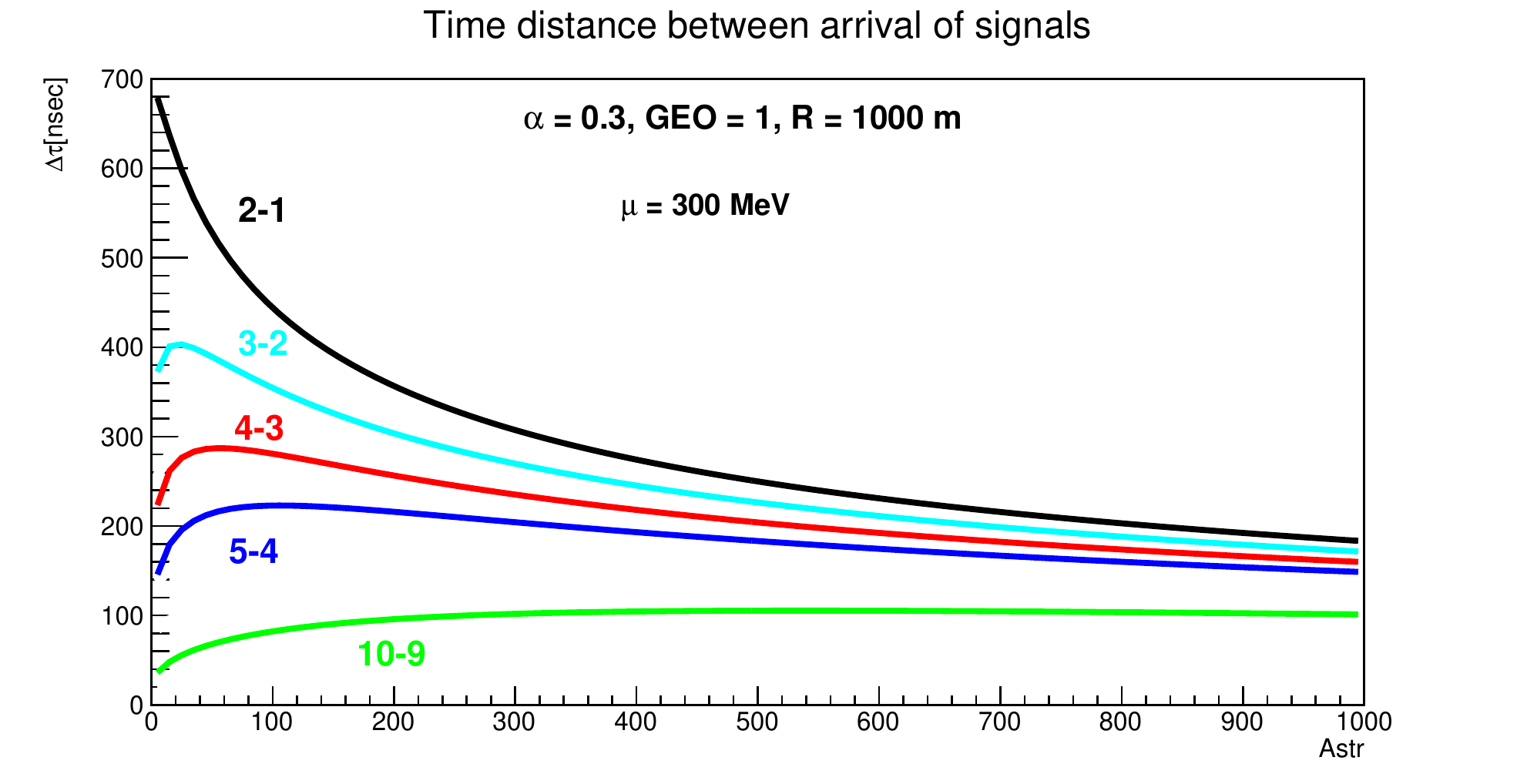}
\includegraphics[width=0.48\linewidth]{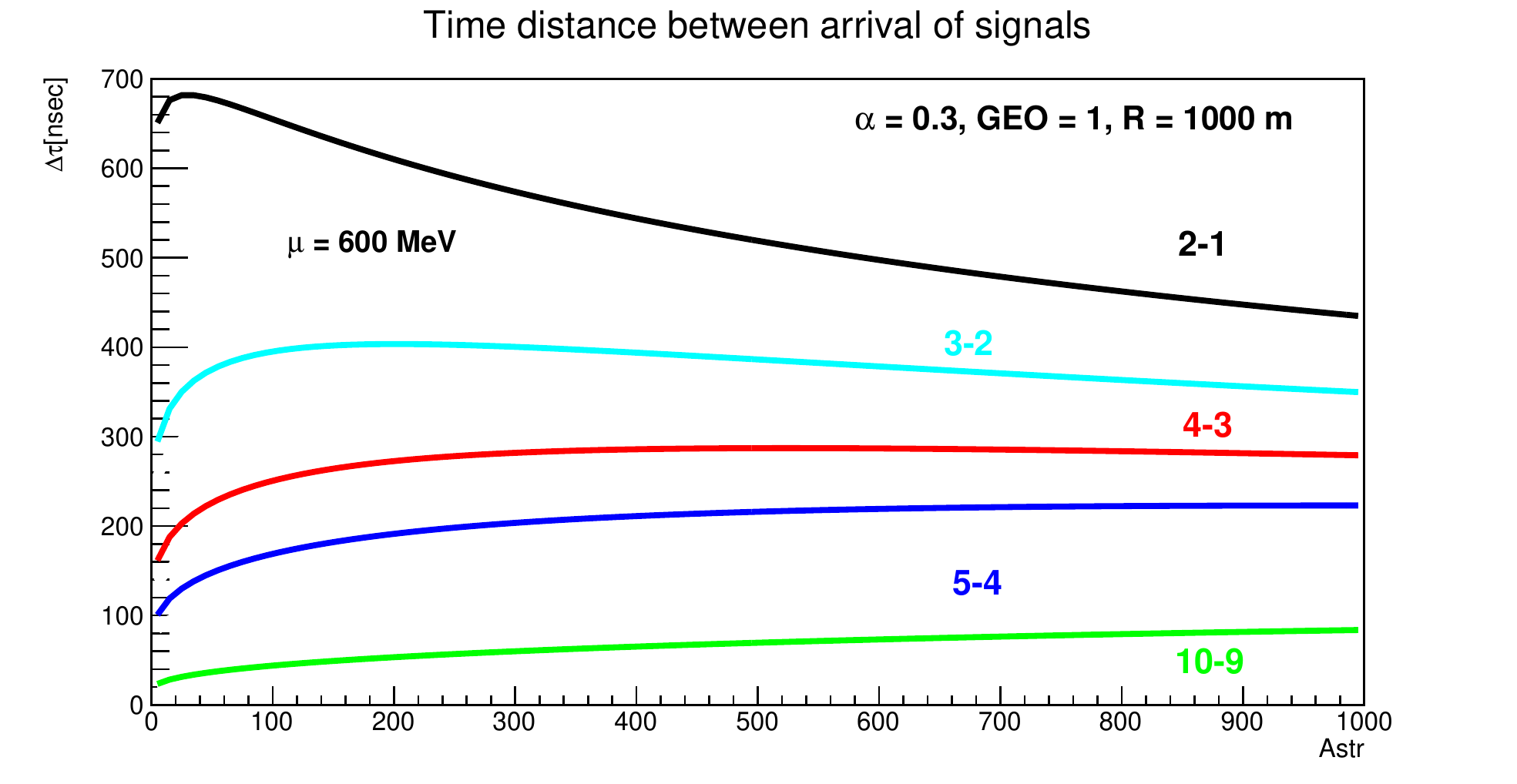}
\includegraphics[width=0.48\linewidth]{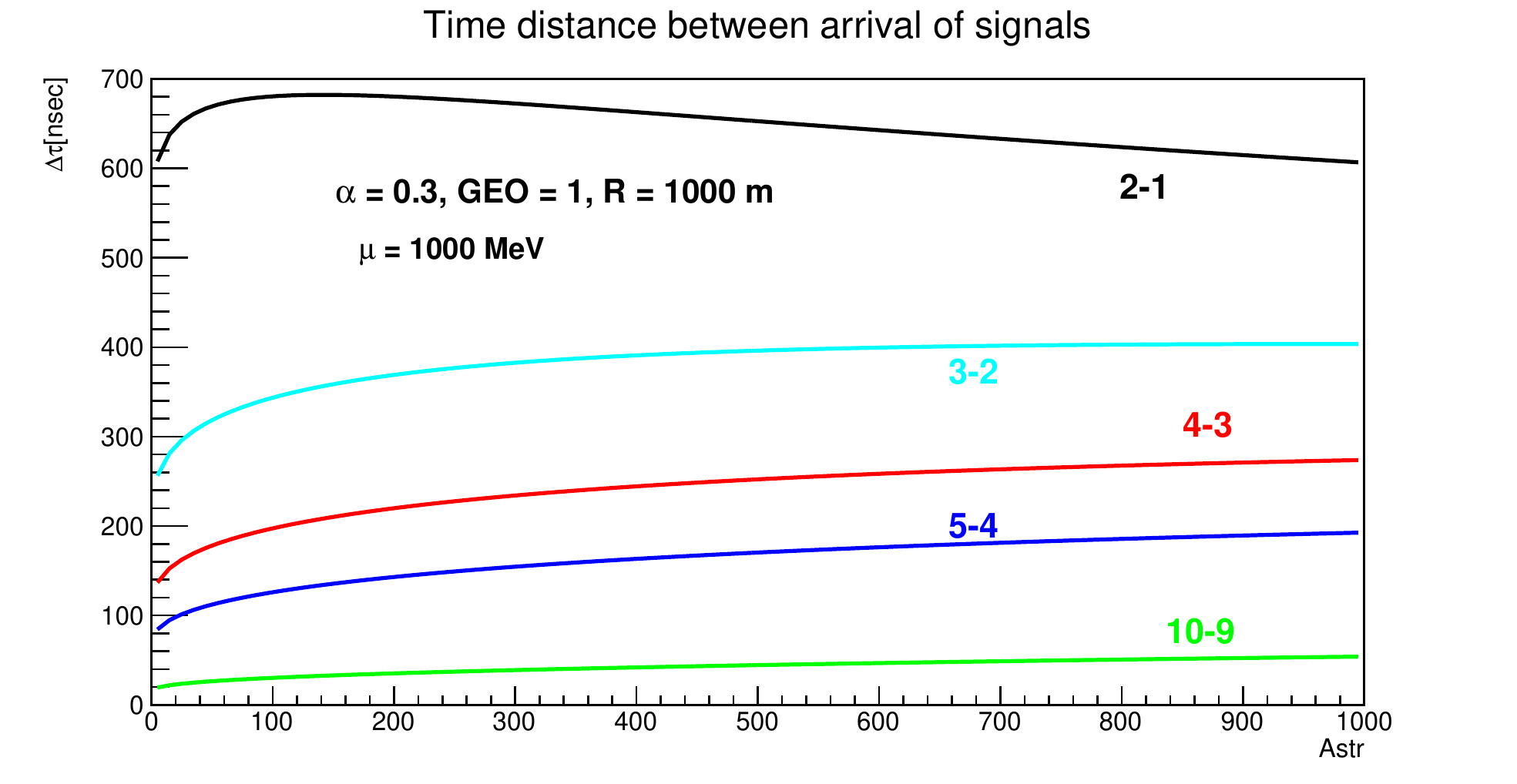}
\caption{ $\Delta\tau$ distances between signals coming from strangelet
interactions at the heights of 2$\lambda_{str}$ and 1$\lambda_{str}$, 3$\lambda_{str}$ and  2$\lambda_{str}$,
 4$\lambda_{str}$ and 3$\lambda_{str}$,  5$\lambda_{str}$ and  4$\lambda_{str}$, 10$\lambda_{str}$ and 
 9$\lambda_{str}$  above the detector level. 
Calculations have been performed  for $\alpha_{s}$ = 0.3,  GEO = 1 and $R$ = 1000 m from the EAS axis,
as a function of a strangelet mass number $Astr$. Two upper pictures have been performed for $\mu$ = 300 MeV
and 600 MeV respectively and the lower one by assuming $\mu$ = 1000 MeV.}
\label{fig:Delta_tau_alfa03_geo1}
\end{center}
\vspace*{-0.8cm}
\end{figure}

\begin{figure}[t]
\vspace*{-2.cm}
\begin{center}
\includegraphics[width=0.48\linewidth]{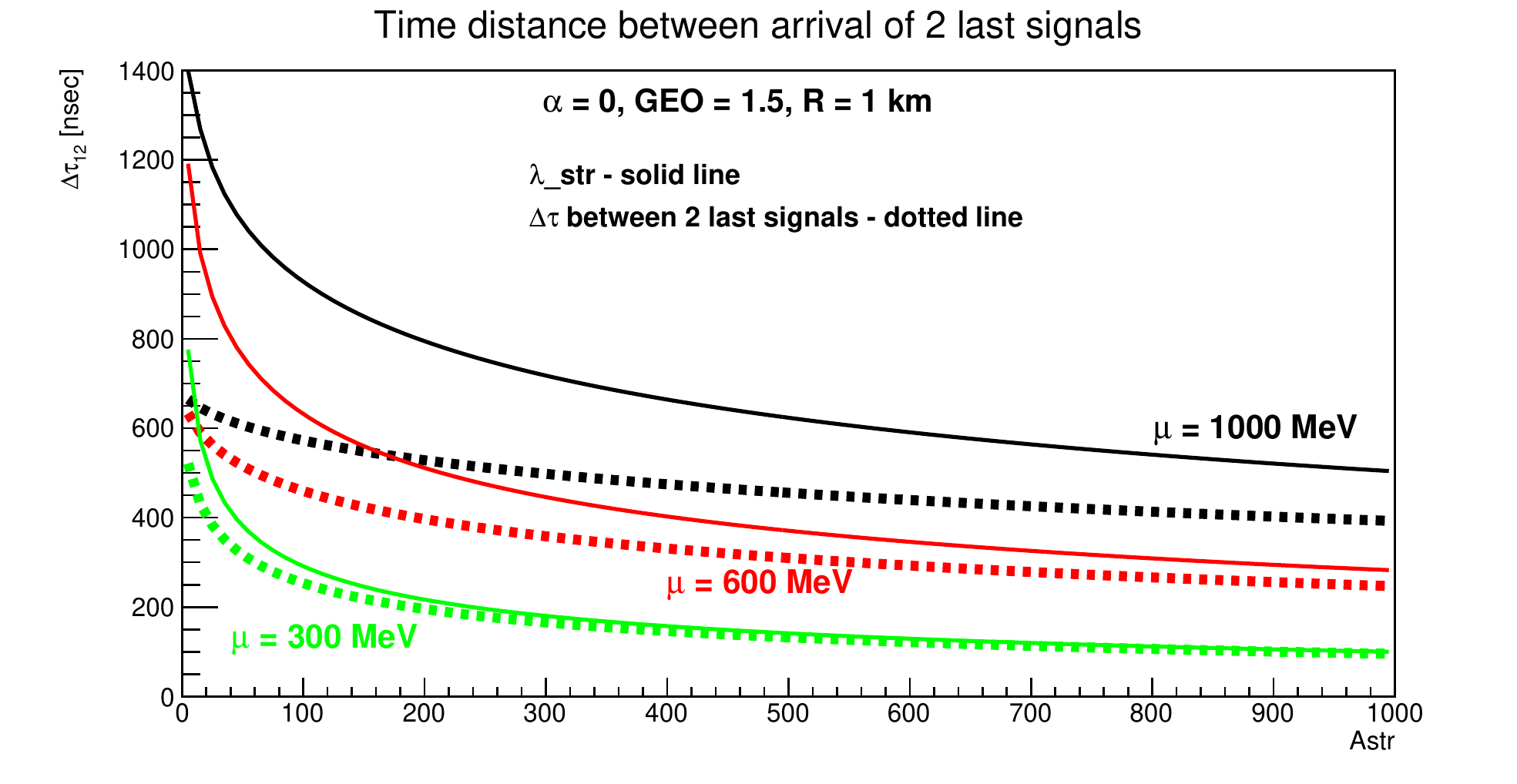}
\includegraphics[width=0.48\linewidth]{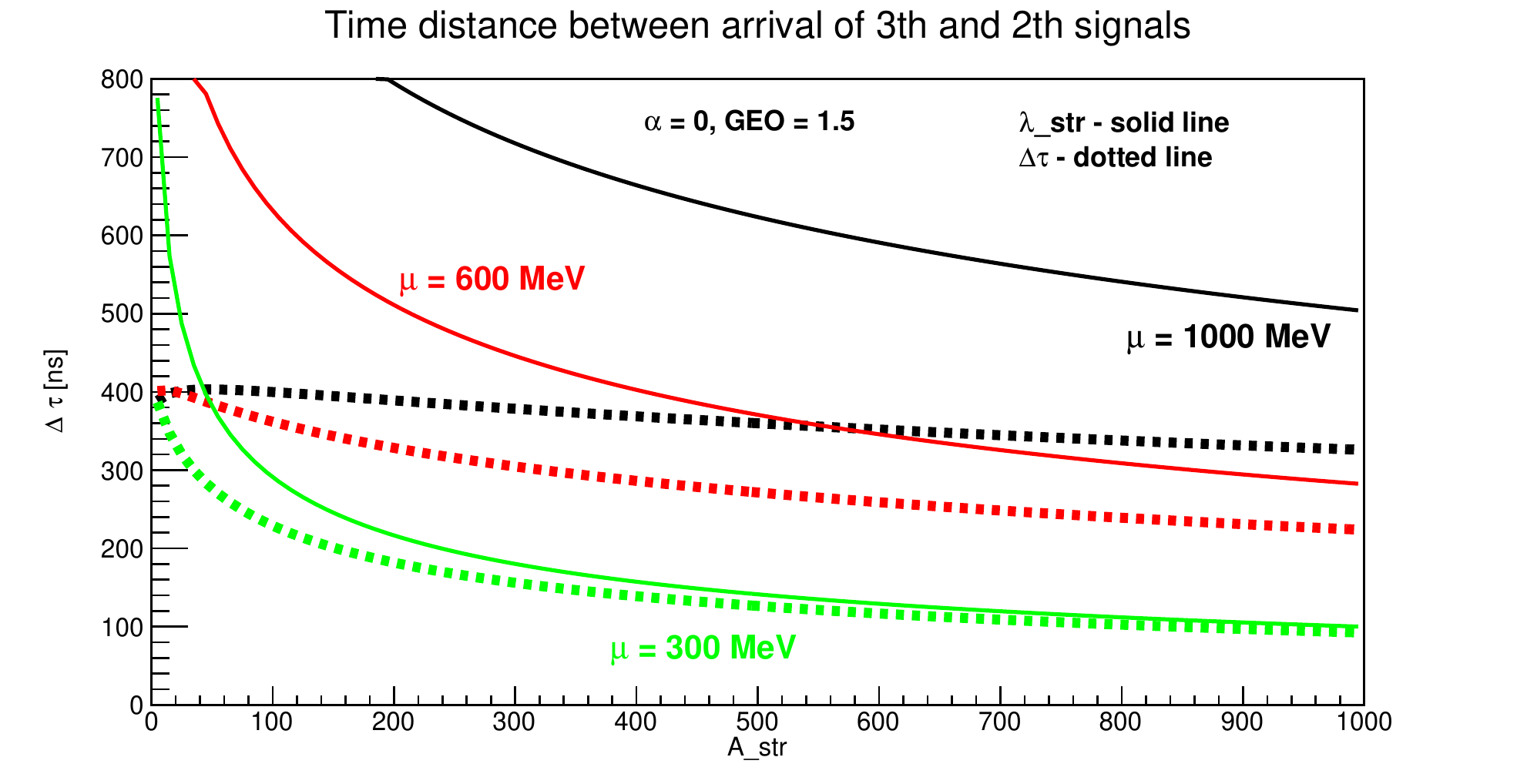}
\hspace*{4mm}\includegraphics[width=0.48\linewidth]{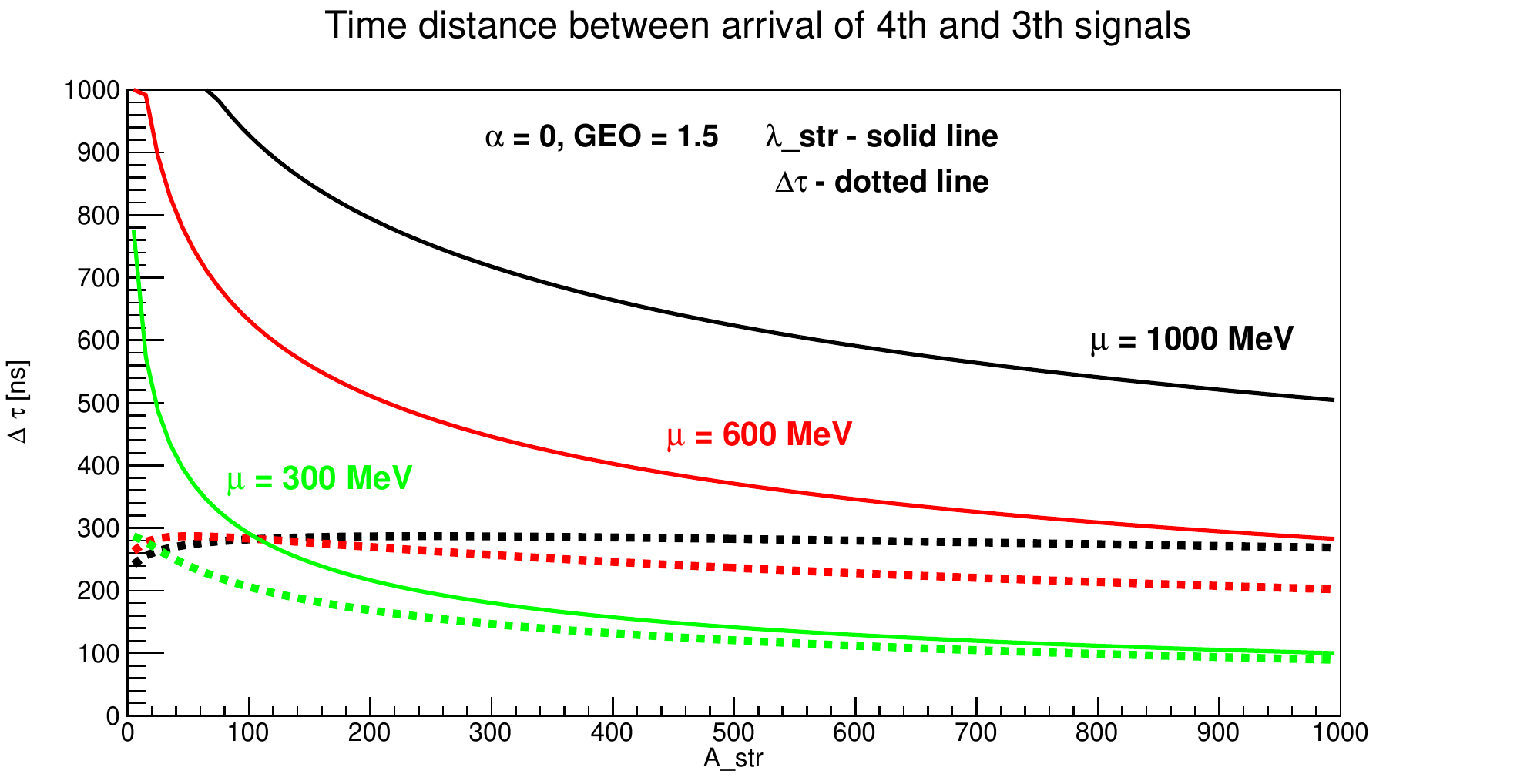}
\hspace*{-3mm}\includegraphics[width=0.48\linewidth]{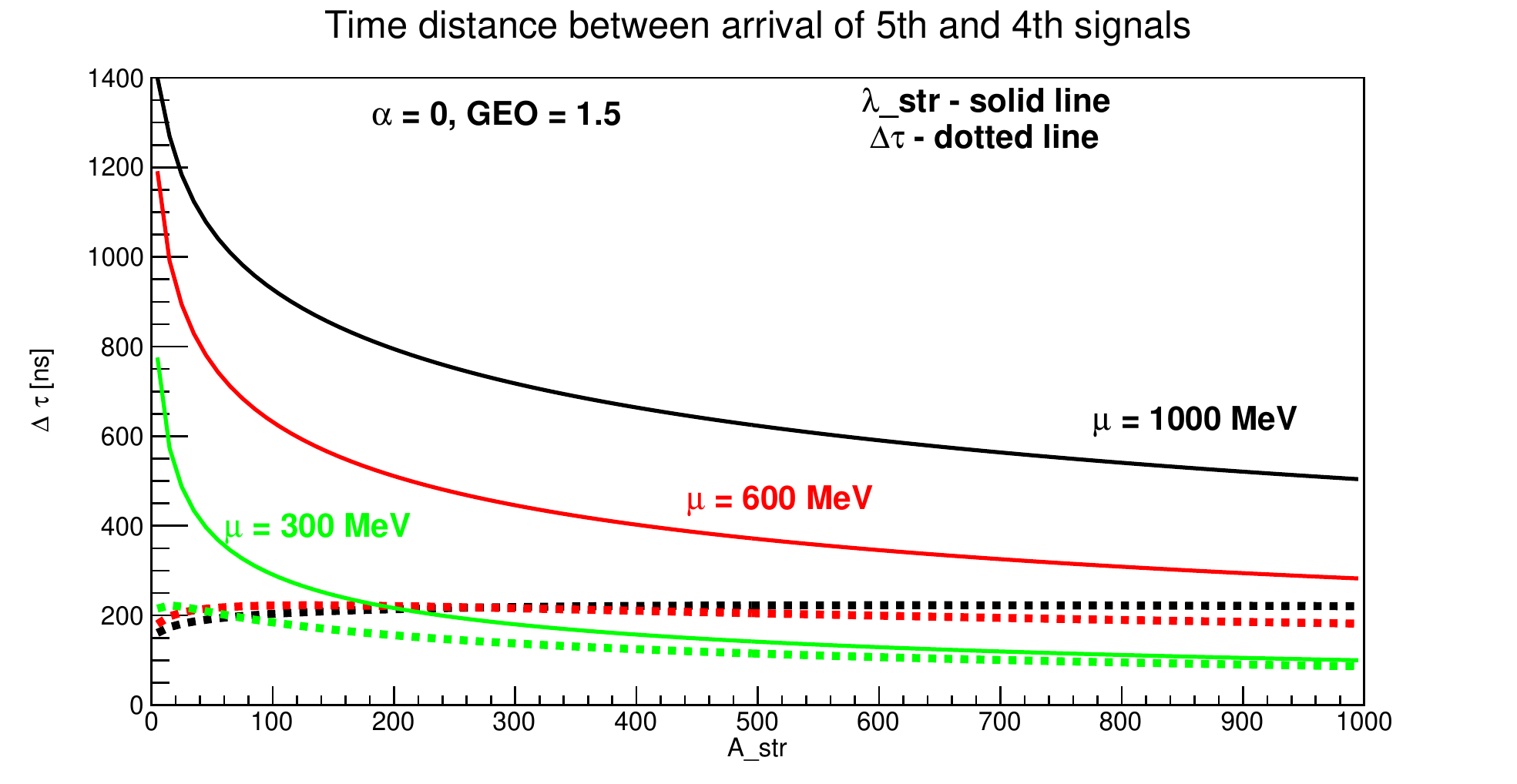}
\includegraphics[width=0.48\linewidth]{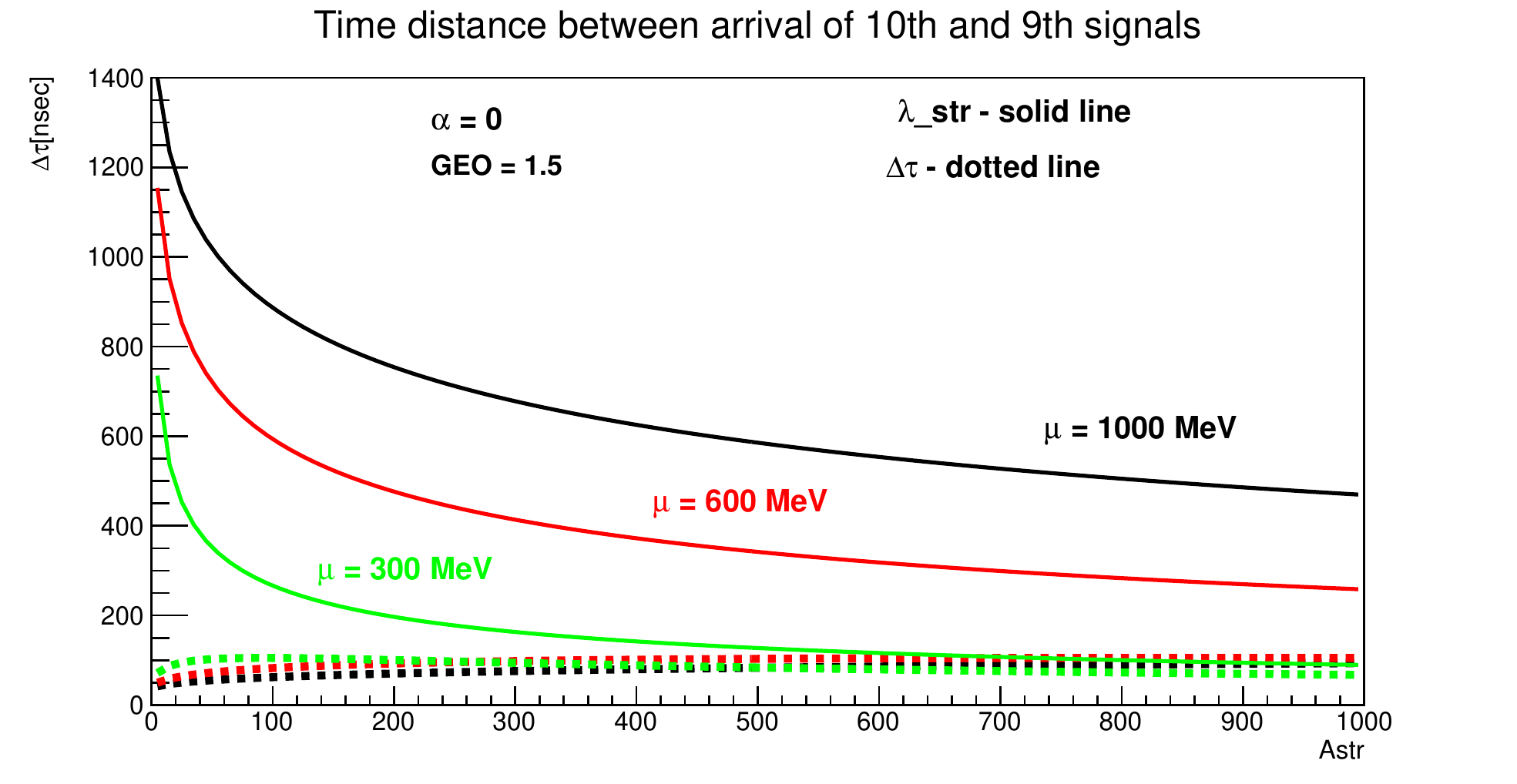}
\caption{ $\Delta\tau$ distances between signals coming from strangelet
interactions at the heights of 2$\lambda_{str}$ and 1$\lambda_{str}$, 3$\lambda_{str}$ and  2$\lambda_{str}$,
 4$\lambda_{str}$ and 3$\lambda_{str}$,  5$\lambda_{str}$ and  4$\lambda_{str}$, 10$\lambda_{str}$ and 
 9$\lambda_{str}$  above the detector level (dotted lines). 
Calculations have been performed  for $\alpha_{s}$ = 0,  GEO = 1.5 and $R$ = 1000 m from the EAS axis,
as a function of a strangelet mass number $Astr$. Comparison 
with the calculated  $\lambda_{str}$ (solid lines) is shown.}
\label{fig:Delta_tau_alfa0_geo1.5}
\end{center}
\vspace*{-1cm}
\end{figure}

Figs.~\ref{fig:Delta_tau_alfa03_geo1}
 and ~\ref{fig:Delta_tau_alfa0_geo1.5} summarized the predicted behaviour of  $\Delta\tau_{n(n-1)}$ distances.

Fig.~\ref{fig:Delta_tau_alfa03_geo1} shows the time distances  between signals 
coming from strangelet
interactions at the heights of 2$\lambda_{str}$ and 1$\lambda_{str}$, 3$\lambda_{str}$ and  2$\lambda_{str}$,
 4$\lambda_{str}$ and 3$\lambda_{str}$,  5$\lambda_{str}$ and  4$\lambda_{str}$, 10$\lambda_{str}$ and 
 9$\lambda_{str}$  above the detector level, as a function of $A_{str}$. 
Calculations have been performed  for $\alpha_{s}$ = 0.3,  GEO = 1 and $R$ = 1000 m from the EAS axis.
 Two upper pictures have been performed for $\mu$ = 300 MeV
and 600 MeV respectively and the lower one for $\mu$ = 1000 MeV. For all $\mu$ the increase 
of $\Delta\tau$ with  a decrease of as well as $Astr$ and $\Delta\tau_{n(n-1)}$ indices is predicted. 
Keeping in mind that $\Delta\tau$ indices mean the height of  the possible strangelet interactions
 (expressed in a number  $\lambda_{str}$), it directly proofs the increase  
of $\Delta\tau$ with the time arrival of signals  to the detector.

Fig.~\ref{fig:Delta_tau_alfa0_geo1.5} 
 shows  $\Delta\tau$ distances between signals coming from strangelet
interactions at several different heights  above the detector level, 
calculated for $\alpha_{s}$ = 0, GEO = 1.5, three  $\mu$ values (300, 600 and 1000 MeV) and 
 $R$ = 1000 m from the EAS axis. Comparison 
with the calculated  $\lambda_{str}$ is also presented.
 Also  this figure supports the statement about {\bf the increase of $\Delta\tau$ values with the
 arrival time of the successive signals to the detector and confirms the weak dependence of this effect
on the assumed values of  parameters.}
 It is also important to note that {\bf for such long distances of the detector
 from the EAS axis (R = 1000 m)  
  $\Delta\tau$ values lie very close to (or between) the calculated $\lambda_{str}$ curves}. 

 Summarizing our investigation of  $\Delta\tau$  behaviour  we can conclude that {\bf the proposed
scenario predicts the experimentally observed behaviour of unusual MME events.}  
 {\bf $\Delta\tau$ are longer for
the signals coming from interactions closer to the detector level than these coming from the high altitudes, i.e.
$\Delta\tau_{21} > \Delta\tau_{32} > \Delta\tau_{43} >...\Delta\tau_{89}$ etc.} 
      
\subsubsection{Correspondence to the detected events}

We have checked the correspondence of the proposed explanation to the experimental data  
comparing the  characteristics of two MME events described in the section 2.2 with the  
predictions of our model. The comparison  is illustrated in 
Figures 13-17.

Our purpose was as well as  to  investigate the general  
correspondence of the experimental data to the theoretical predictions  
and, in particular,  to study a possible connection between 
the multimodal pictures obtained from different detection stations.

\begin{center}
\begin{large}
{\bf  5-peak MME event published in \cite{Beisembaev_5peaks}}
\end{large}
\end{center}

The event has been already described in the section 2.2. It is shown in  Fig.~\ref{fig:MME_5peaks}
  and its characteristics are summarized in the  Table~\ref{tab-5peaks}.

We have compared the experimental $\Delta\tau$ values with the calculated ones for two series of parameters.

\begin{figure}
\vspace*{-1.5cm}
\begin{center}
\includegraphics[width=0.49\linewidth]{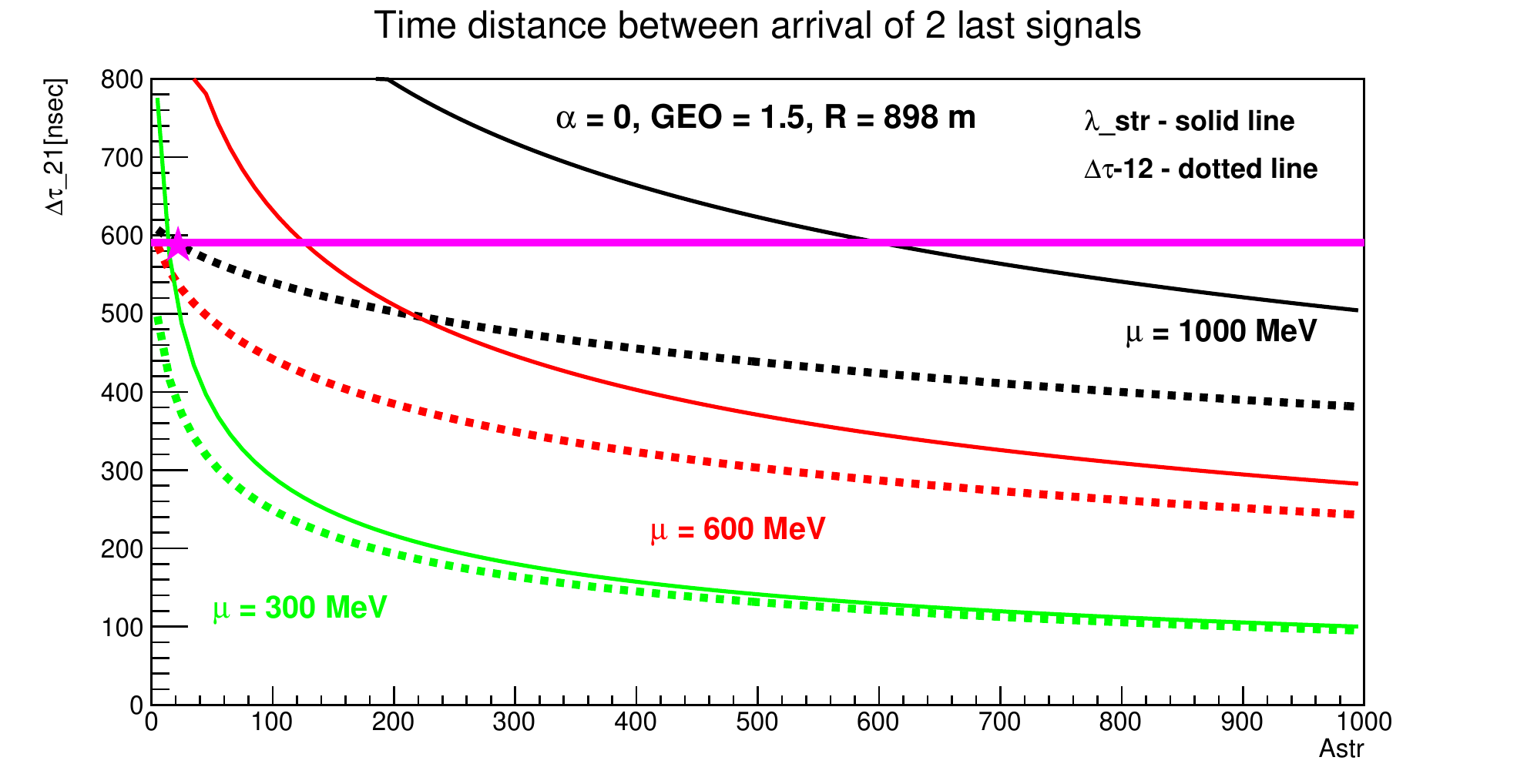}
\includegraphics[width=0.49\linewidth]{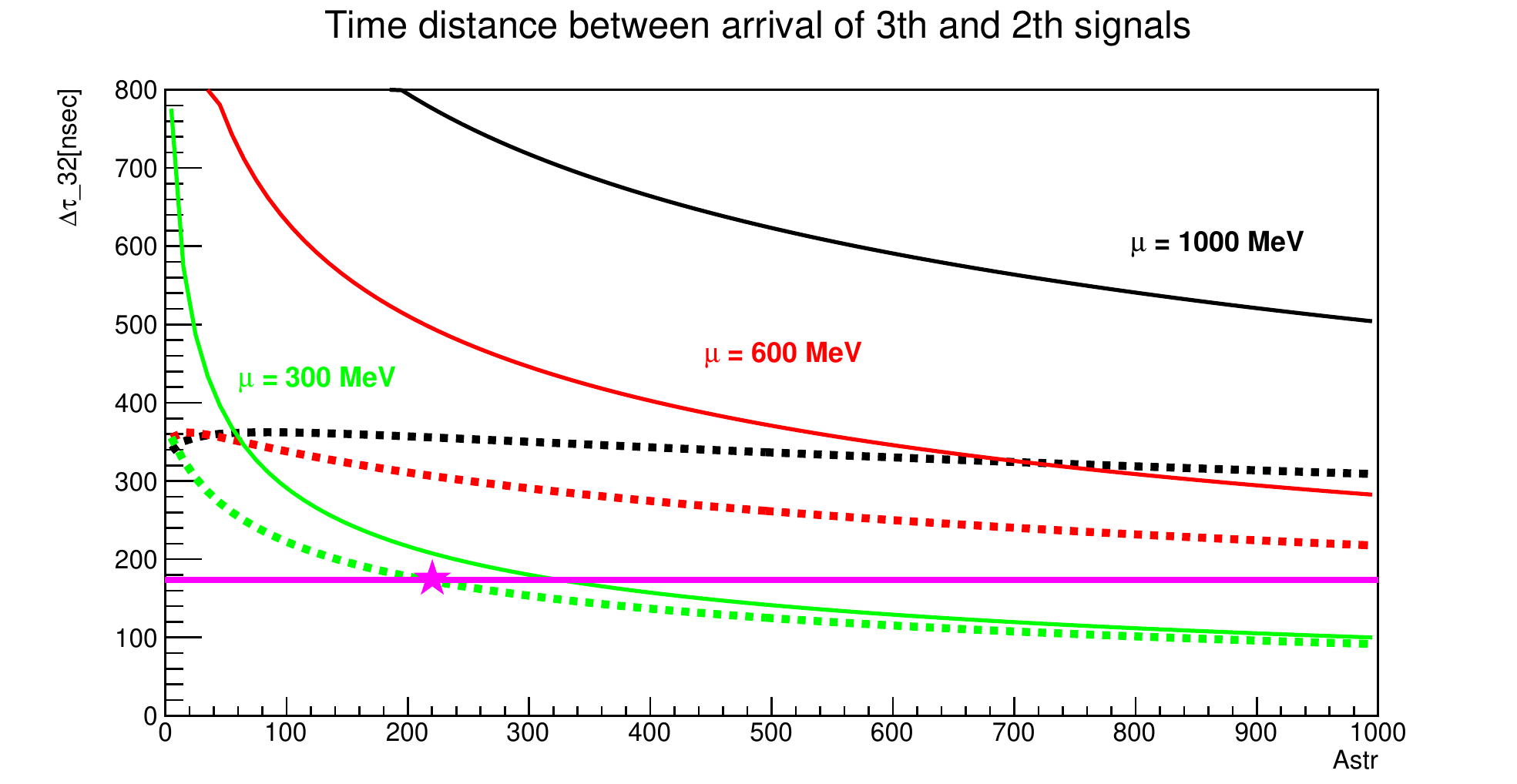}
\includegraphics[width=0.49\linewidth]{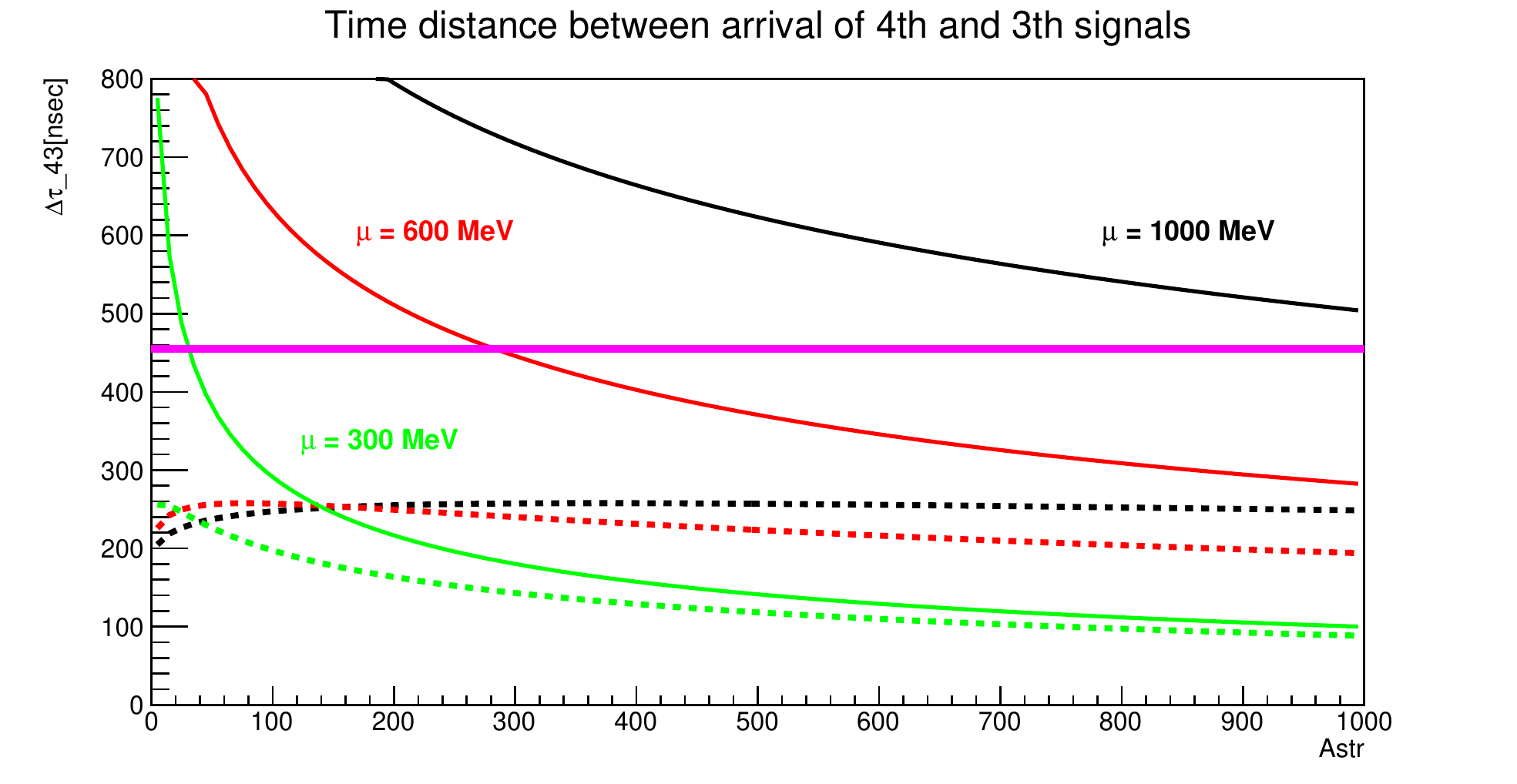}
\includegraphics[width=0.49\linewidth]{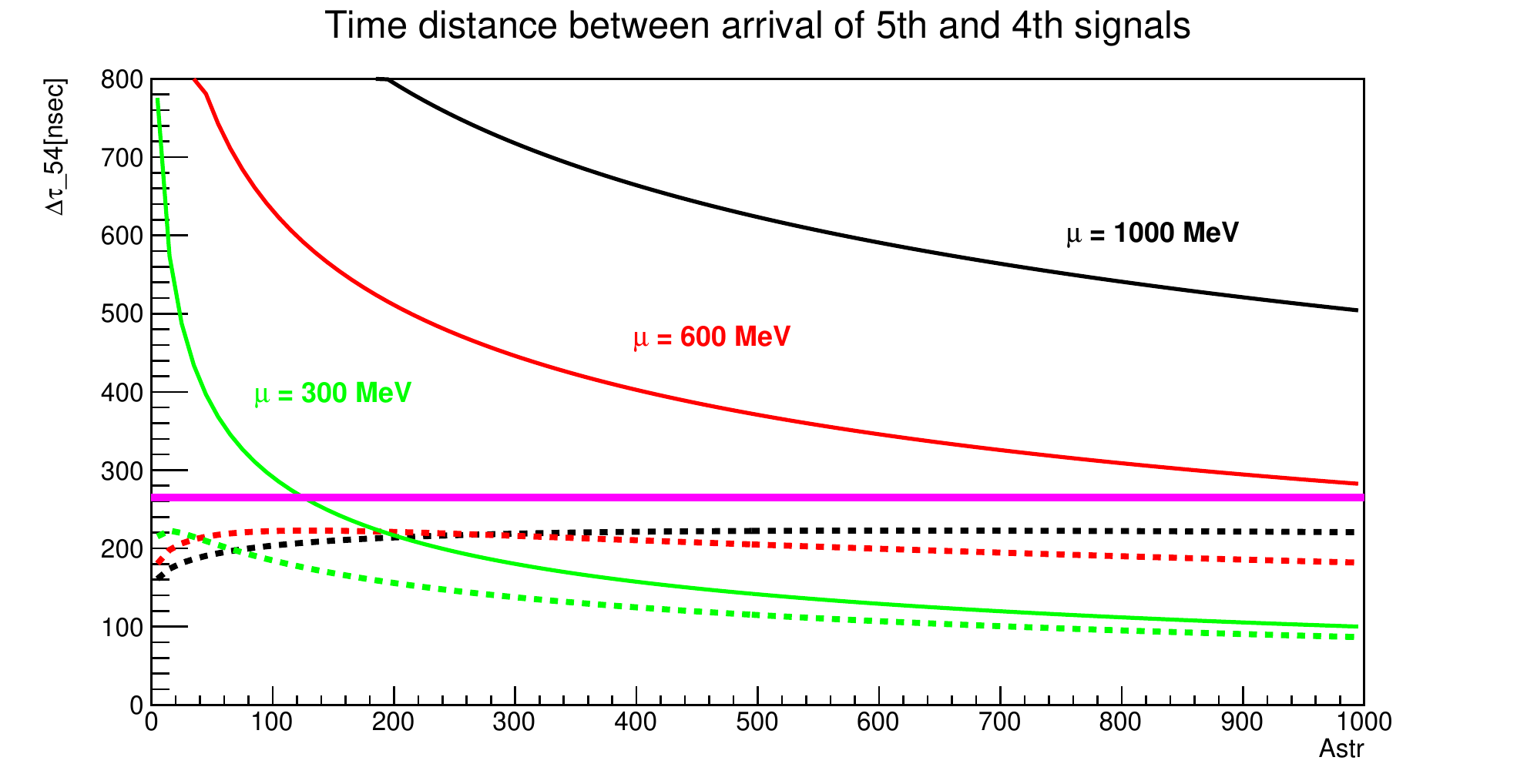}
\caption{ $\Delta\tau$ distances between signals coming from strangelet
interactions at the altitudes of 2$\lambda_{str}$ and 1$\lambda_{str}$ ,  
3$\lambda_{str}$ and 2$\lambda_{str}$ , 4$\lambda_{str}$ and 3$\lambda_{str}$ , 
 5$\lambda_{str}$ and
 4$\lambda_{str}$  above the detector level and
calculated for $R$ = 898 m from the EAS axis, as a function of a strangelet mass number $Astr$ are shown by the 
dotted lines.  
Calculated $\lambda_{str}$ are shown by the solid lines. Colours: black, red and green 
 correspond to $\mu$ = 1000, 600 and 300 MeV respectively. 
$\alpha_{s}$ = 0 and GEO = 1.5 have been assumed. Experimental values of $\Delta\tau$ for
5-peak MME event \cite{Beisembaev_5peaks}, have been shown by the  violet straight lines. Intersections of  
the experimental  $\Delta\tau_{21}$ and $\Delta\tau_{32}$ values with the calculated 
$\Delta\tau$ curves are marked by stars.} 
\label{fig:5peaks_GEO15_alfa0}
\end{center}
\vspace*{-0.4cm}
\end{figure}

Fig.~\ref{fig:5peaks_GEO15_alfa0} shows 
$\Delta\tau$ distances between signals coming from strangelet
interactions at the heights of 2$\lambda_{str}$ and 1$\lambda_{str}$,   
3$\lambda_{str}$ and 2$\lambda_{str}$, 4$\lambda_{str}$ and 3$\lambda_{str}$, 5$\lambda_{str}$ and
 4$\lambda_{str}$ above the detector level, 
calculated for three values of $\mu$ (300, 600 and 1000 MeV) and for $R = 898$ m from the EAS axis.   
Calculated  $\lambda_{str}$ are also presented. 
$\alpha_{s}$ = 0 and GEO = 1.5 have been assumed. Experimental values of $\Delta\tau$ for
5-peak MME event \cite{Beisembaev_5peaks}, have been shown by  the violet straight lines. Because 
of fluctuations we can not, of course, 
expect the direct correspondence between the features of single experimental events and the theoretical 
predictions.   It is seen, however,  that for this event
{\bf the experimental $\Delta\tau$ values are within the calculated $\lambda_{str}$ curves}.  
 Even more,  intersections of
the violet lines with the calculated  $\Delta\tau_{21}$ at $Astr \approx 20$ and $\Delta\tau_{32}$ at 
$Astr \sim 230$ (marked by stars) suggest that the event could arise from the interaction of a  strangelet 
with $Astr \ge 230$
(and $\mu \ge 300$ MeV) at the altitude of $\sim 3 \lambda_{str}$ resulting into a 
strangelet characterized  by $Astr \le 20$ and $\mu \le$ 600 MeV which interacted at the altitude of
  $\sim 2 \lambda_{str}$ above the detector level. The observed increase of
the strangelet mass number $Astr$ with the altitude, i.e.   
 a decrease of $Astr$ with the
number of the successive collisions is in accordance with the proposed scenario.

The experimental  $\Delta\tau_{43}$ and $\Delta\tau_{54}$ values are, however, far from the calculated 
$\Delta\tau$ values. There are several possible explanations of such observation. High  $\Delta\tau_{43}$  
and $\Delta\tau_{54}$ values could be caused simply by fluctuatios, by a very 
massive ($Astr > 1000$) strangelet interacting  at the altitudes
of $\sim 4 \lambda_{str}$ and $\sim 5 \lambda_{str}$ above the detector level
  or 
 by the additional, very weak interactions, with the signal below the detection threshold, at the altitude 
 $\sim (3-5) \lambda_{str}$.

\begin{figure}
\begin{center}
\includegraphics[width=1.\linewidth]{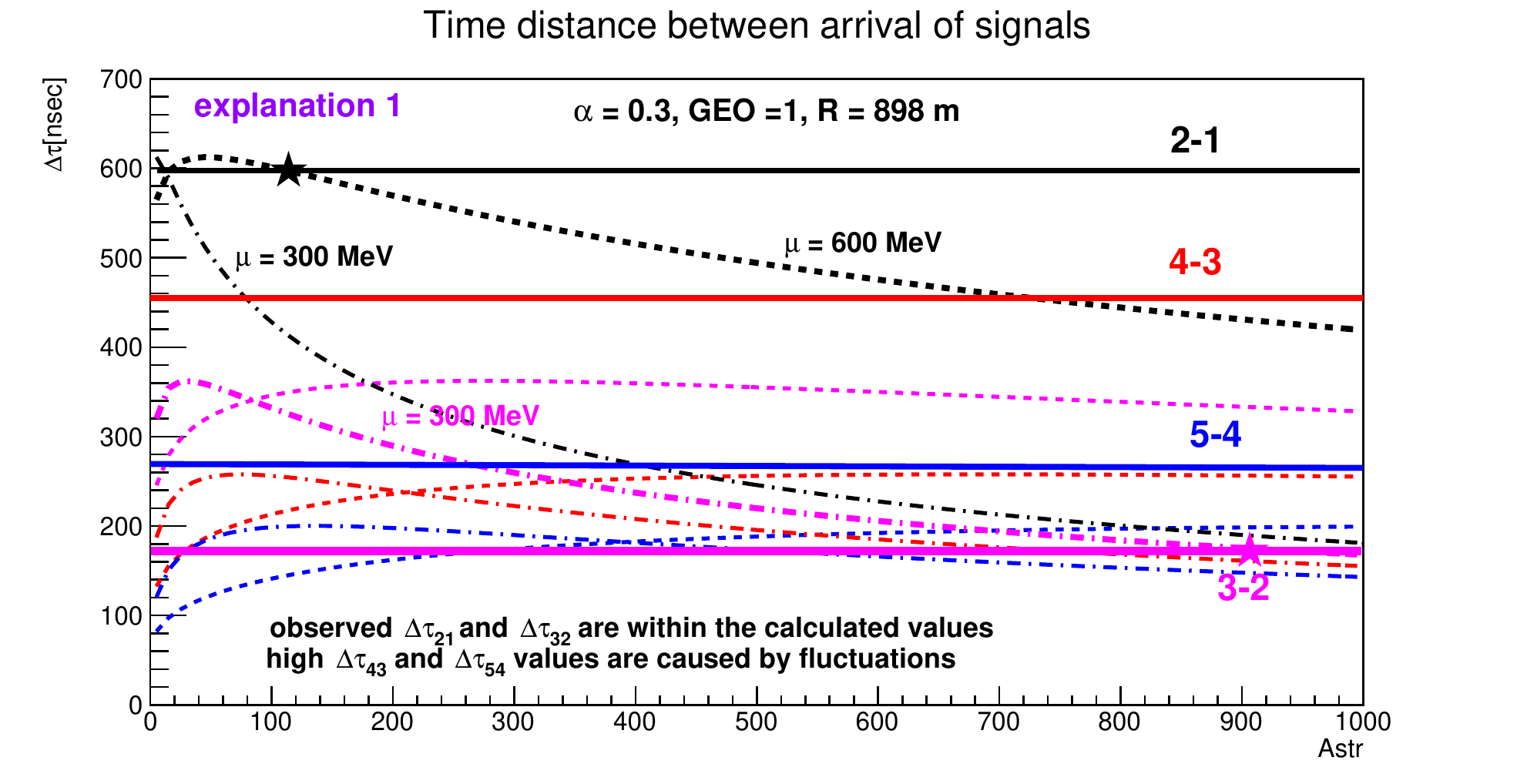}
\includegraphics[width=1.\linewidth]{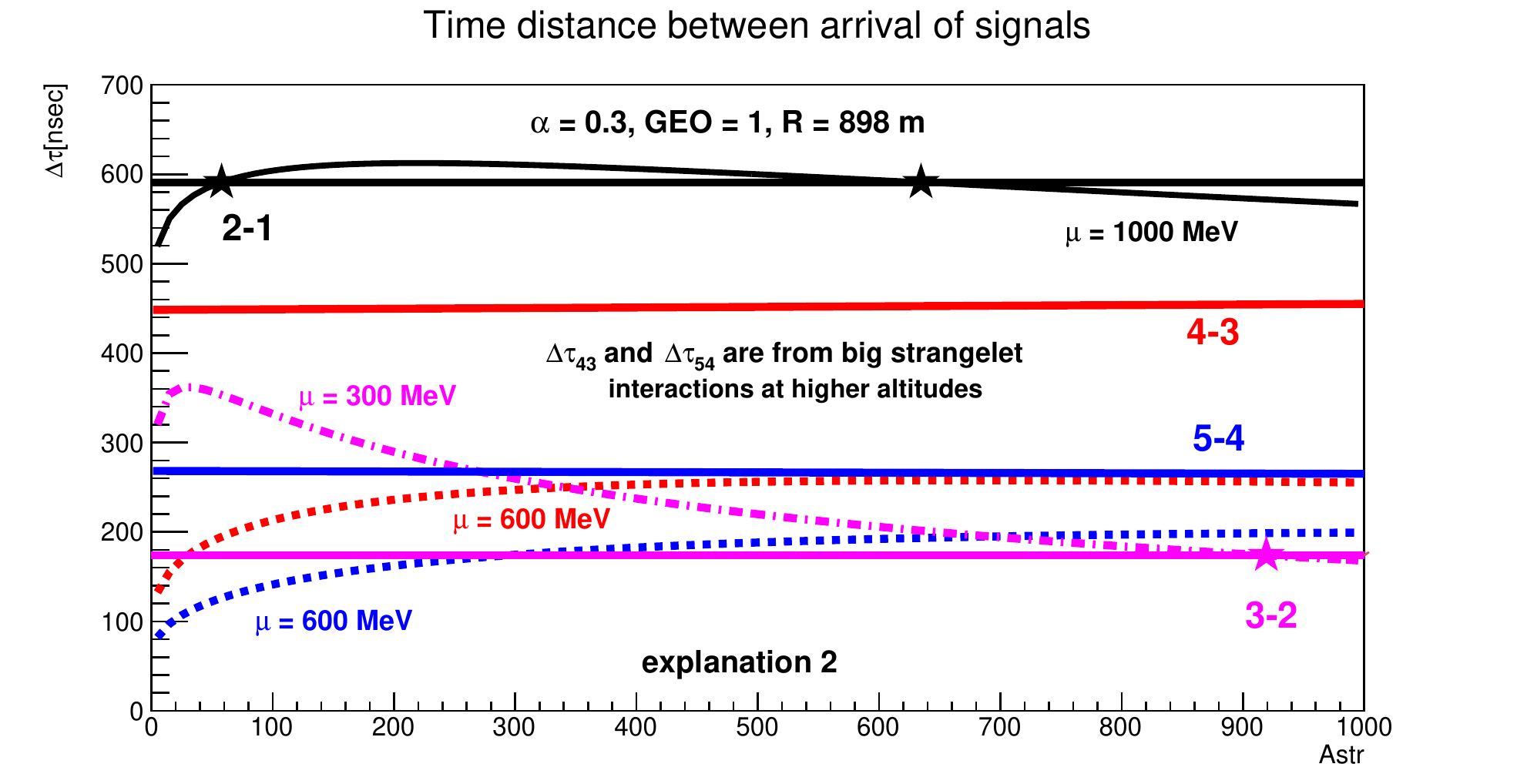}
\caption{$\Delta\tau$ distances between signals coming from strangelet
interactions at the heights of 2$\lambda_{str}$ and 1$\lambda_{str}$,  
3$\lambda_{str}$  and 2$\lambda_{str}$,  4$\lambda_{str}$ and 3$\lambda_{str}$  
or  5$\lambda_{str}$ and
 4$\lambda_{str}$  above the detector level and
calculated for $R$ = 898 m from the EAS axis, as a function of a strangelet mass number $Astr$.
$\alpha_{s}$ = 0.3, GEO = 1 have been assumed.
In the upper figure only curves for $\mu$ = 300 MeV and 600 MeV have been shown. 
 Experimental  $\Delta\tau$ values for
5-peak event \cite{Beisembaev_5peaks}, have been shown by the straight lines in the same colour as 
the corresponding $\Delta\tau$.}
\label{fig:5peaks_GEO1_alfa03}
\end{center}
\vspace*{-0.4cm}
\end{figure}

Fig.~\ref{fig:5peaks_GEO15_alfa0}  illustrates the possible
explanations of the 5-peak MME event by comparison of the experimental $\Delta\tau$ values with 
these calculated for GEO = 1.5 and $\alpha_{s}$ = 0.
The general interpretation of this figure is consistent with that resulting from Fig.~\ref{fig:5peaks_GEO1_alfa03},
 obtained by assuming GEO = 1 and $\alpha_{s}$ = 0.3.
In both cases the experimental  $\Delta\tau_{21}$  and $\Delta\tau_{32}$ values are 
in agreement with the theoretical expectations.  Both $\Delta\tau_{21}$ values could be explained by the successive 
strangelet
interactions at the heights of 2$\lambda_{str}$ and 1$\lambda_{str}$ above the detector level, and 
 $\Delta\tau_{32}$ values  by the  strangelet
interactions at the heights of 3$\lambda_{str}$ and 2$\lambda_{str}$. Independently on the values of
parameters which we used in calculations, the estimated $Astr$ corresponding to $\Delta\tau_{32}$  is greater than
 $Astr$ for $\Delta\tau_{21}$. The possible estimated $Astr$ values 
 are: $Astr \sim 20 - 120$ for $\Delta\tau_{21}$ and  $\mu \approx$  600-1000 MeV, 
$Astr \approx 230-920$ for $\Delta\tau_{32}$ and $\mu \approx$ 300-600 MeV.

For both series of parameters  the  straight lines  
showing the experimental   $\Delta\tau_{43}$ 
and $\Delta\tau_{54}$ values are above the calculated ones. This observation could be  explained
by fluctuations or the appearance of the additional undetected strangelet interactions.

Summarizing, we can say  that the experimental characteristics of this event support the proposed
scenario.  The theoretically predicted  
  decrease of $\Delta\tau$ with
the increase of as well as $Astr$ and the height of a strangelet interaction is observed.

\begin{center}
\begin{large}
{\bf  MME event published in \cite{peaks-3-8-4-6}}
\end{large}
\end{center}

As it was described in the section 2.2 (see Fig.~\ref{fig:MME-4-6}),  the multimodal pattern 
of this event was observed  
 in several stations. The most apparently seen peak structures have been detected in stations: 
8, 6 and 4. 
    
\begin{figure}
\vspace*{-2cm}
\begin{center}
\includegraphics[width=1. \linewidth]{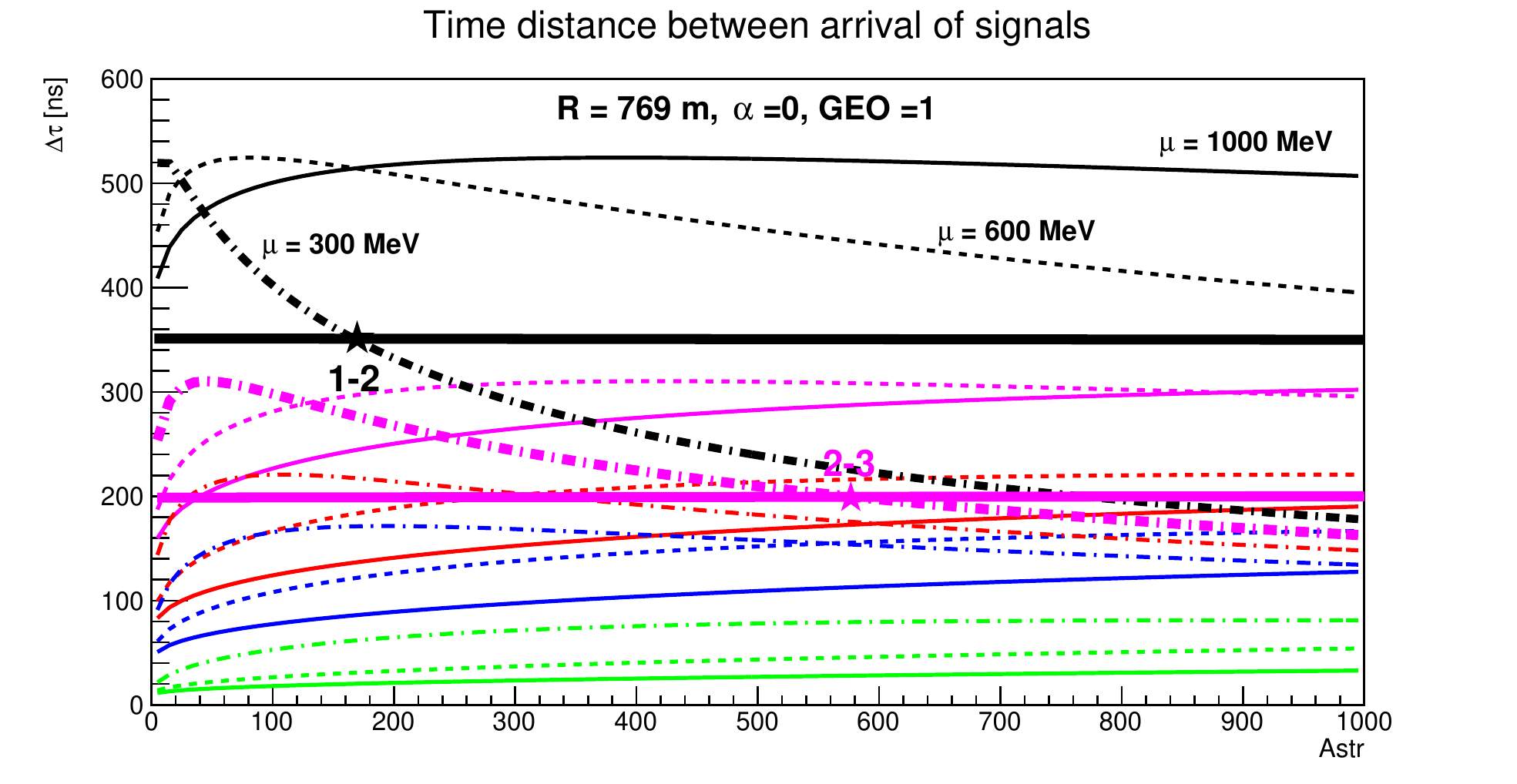}
\caption{ $\Delta\tau$ distances between signals coming from strangelet
interactions at the heights  of 2$\lambda_{str}$ and 1$\lambda_{str}$ (black),  
3$\lambda_{str}$  and 2$\lambda_{str}$ (violet),  4$\lambda_{str}$ and 3$\lambda_{str}$ (red), 
  5$\lambda_{str}$ and 4$\lambda_{str}$ (blue),
 10 $\lambda_{str}$ and 9$\lambda_{str}$ (green) above the detector level, 
as a function of a strangelet mass number $Astr$. 
Calculations have been performed  for $\alpha_{s}$ = 0,  GEO = 1, $\mu$ = 300, 600 and 1000 MeV,
 and $R$ = 769 m from the EAS axis.
Experimental  $\Delta\tau$ values  are shown by the straight lines with te same colour as 
the corresponding $\Delta\tau$ curve. Intersections of  curves are marked by  stars.} 
\label{fig:R769}
\end{center}
\vspace*{-0.4cm}
\end{figure}

The possible explanation of peaks detected by the station 8, located at the distance $R \approx$ 769 m from
the event axis is shown in Fig.~\ref{fig:R769}.
Calculated  $\Delta\tau$ distances between signals coming from strangelet
interactions at several different heights above the detector level, 
as a function of a strangelet mass number $Astr$ are shown.
Calculations have been performed  for $ \alpha_{s}$ = 0,  GEO = 1, $\mu$ = 300, 600 and 1000 MeV,
 and $R$ = 769 m from the EAS axis.
Experimental  $\Delta\tau$ values  are shown by the straight lines of the same colour as 
the corresponding $\Delta\tau$ curves. Intersections of the experimental lines with the calculated 
curves are marked by  stars.

Assuming the values of distances between peaks
 $\Delta\tau_{21} \simeq$ 200 ns and $\Delta\tau_{32} \simeq$ 350 ns  we 
 find that this event could be explained by the following scenario:
a strangelet  with 
Astr $\ge$ 570 and $\mu \ge$ 300 MeV interacts at the altitude $H \approx 3\lambda_{str}$ 
and the new born strangelet with Astr $\ge$ 180 and $\mu \sim$ 300 MeV interacts at 
the altitude $H \approx 2\lambda_{str}$. 

\begin{figure}
\vspace*{-2cm}
\begin{center}
\includegraphics[width=1. \linewidth]{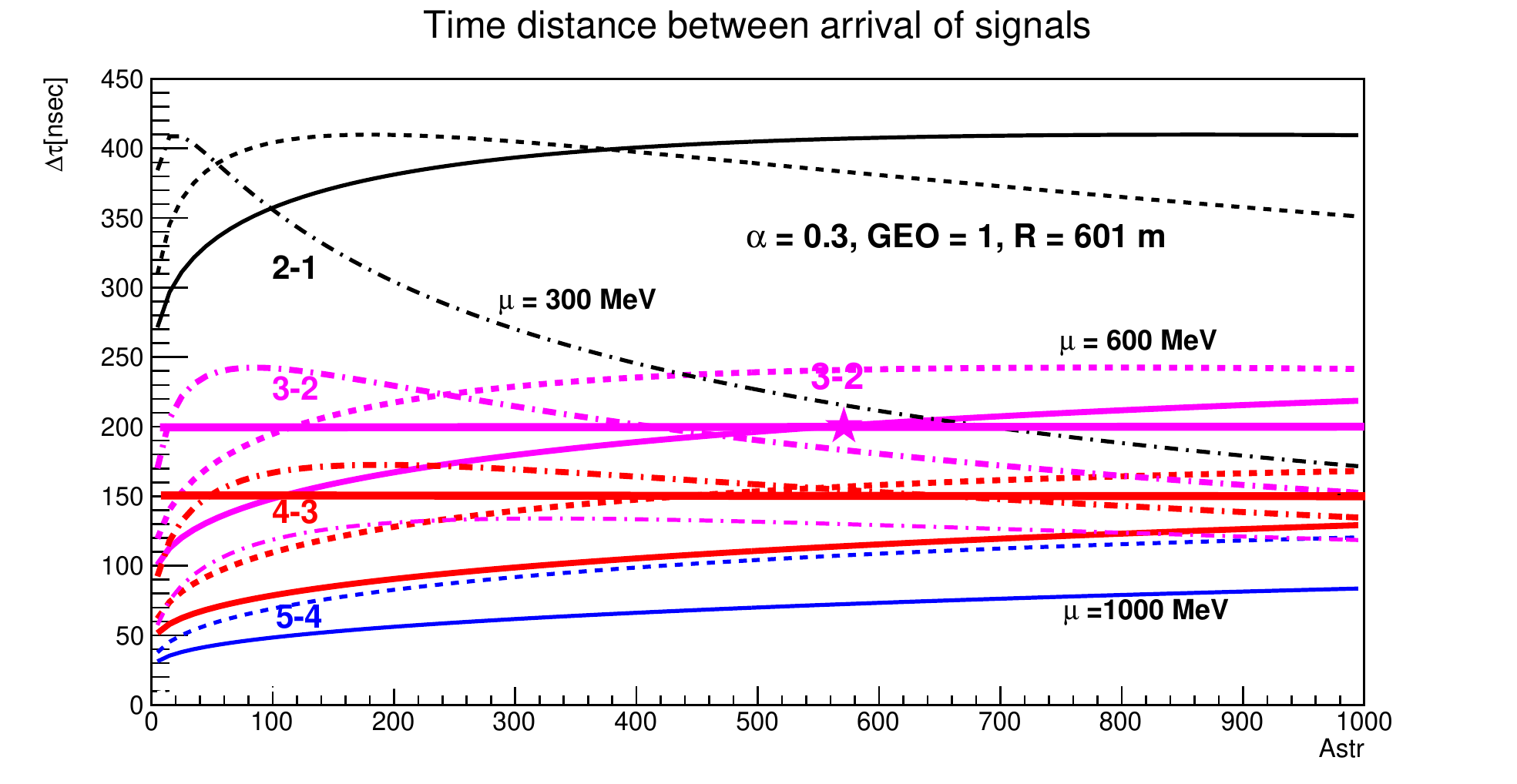}
\caption{ $\Delta\tau$ distances between signals coming from strangelet
interactions at the heights of 5$\lambda_{str}$ and 4$\lambda_{str}$, 4$\lambda_{str}$ and  3$\lambda_{str}$
3$\lambda_{str}$ and 2$\lambda_{str}$, 2$\lambda_{str}$ and  1$\lambda_{str}$,
 above the detector level, as a function of a strangelet mass number $Astr$. 
Calculation have been performed  for $\alpha_{s}$ = 0,  GEO = 1, $\mu$ = 300, 600 and 1000 MeV,
 and $R$ = 601 m from the EAS axis.
Experimental  $\Delta\tau$ values are shown by the straight lines with te same colour as 
the corresponding $\Delta\tau$ curves. Intersections of  curves are marked by  stars. }
\label{fig:R601_alternative}
\end{center}
\vspace*{-0.4cm}
\end{figure}

Fig.~\ref{fig:R601_alternative} shows possible
 explanation of the MME signals detected by the station no 6.
As the amplitude of the signals is very low  it is possible that the last weak signal was lost
\cite{peaks-3-8-4-6}.
In that case the observed time distances between peaks  correspond to $\Delta\tau_{32} \approx$
200 MeV and $\Delta\tau_{43} \approx$ 150 MeV.

As it is seen from  
Fig.~\ref{fig:R601_alternative} the red straight line showing the experimental value of $\Delta\tau_{32}$ lies
between the corresponding  theoretical curves (for $\mu$ = 300, 600 and 1000 MeV) in the total investigated 
range (0 $<$ Astr $<$ 1000). So, we can only say that the experimental $\Delta\tau_{32}$ value is in agreement 
with the theoretical predictions but it is impossible to precise the value of the interacting strangelet mass
number $Astr$. It should be noted, however, that the theoretical $\Delta\tau_{32}$ line, corresponding to $\mu$ = 1000 MeV, is
the closest to the experimental one. So, assuming  that it gives the best description of the data and knowing  
   the  intersection point of both lines we can expect that the mass number of the strangelet interacting at the altitude 
 of $\sim 3\lambda_{str}$ above the detector level  $Astr \approx 580$. 

The same problem concerns also the interpretation of $\Delta\tau_{43}$ line. The violet straight line showing 
the experimental value of $\Delta\tau_{43}$ lies
between the corresponding  theoretical curves (for $\mu$ = 300, 600 and 1000 MeV) in the total investigated 
range (0 $<$ Astr $<$ 1000). In this case, we are only able to say that the experimental $\Delta\tau_{43}$ 
values are in agreement
with the theoretical predictions and we are not able to estimate how big the strangelet interacting at the altitude 
of $4\lambda_{str}$ could be.

Summarizing, the results obtained in the stations 8 and 6, could be explained by the following 
picture. A big strangelet  interacts at the 
altitude  of $\sim 4\lambda{str}$ above the detector level. The successive interactions of the new born  
 strangelets with 
Astr $\ge$ 580 and  Astr $\ge$ 200 occur at the altitude $H \sim 3\lambda_{str}$ 
and   $\sim 2\lambda_{str}$ respectively
(see Fig.~\ref{fig:R601_alternative}).

\begin{figure}
\vspace{-1cm}
\begin{center}
\includegraphics[width=1. \linewidth]{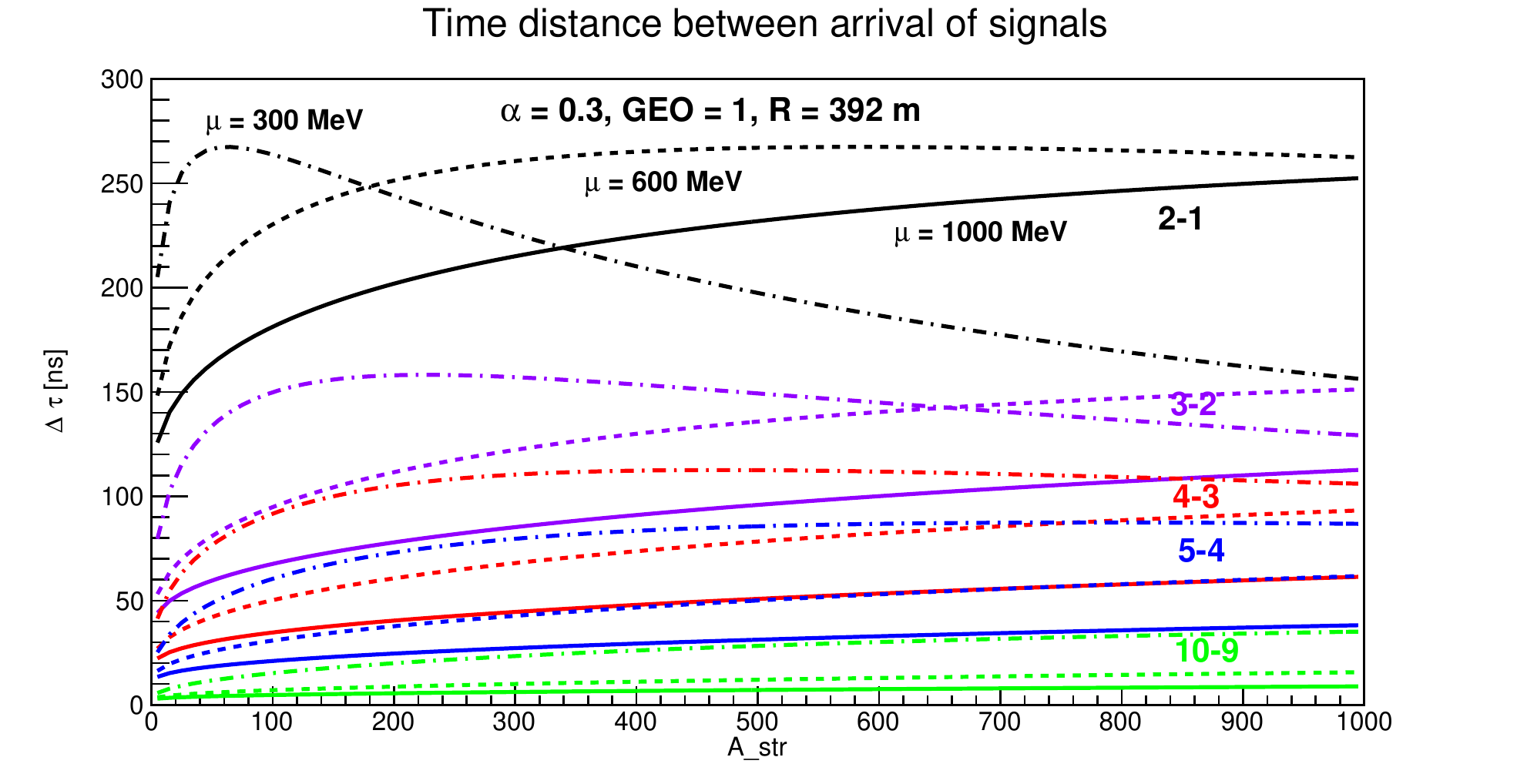}
\caption{ $\Delta\tau$ distances between signals coming from strangelet
interactions at the heights of 10$\lambda_{str}$ and 9$\lambda_{str}$, 
5$\lambda_{str}$ and 4$\lambda_{str}$, 4$\lambda_{str}$ and  3$\lambda_{str}$
3$\lambda_{str}$ and 2$\lambda_{str}$, 2$\lambda_{str}$ and  1$\lambda_{str}$,
 above the detector level, as a function of a strangelet mass number $Astr$. 
Calculation have been performed  for $\alpha_{s}$ = 0.3,  GEO = 1, $\mu$ = 300, 600 and 1000 MeV,
 and $R$ = 392 m from the EAS axis.}
\label{fig:R392}
\end{center}
\vspace*{-0.4cm}
\end{figure}

Fig.~\ref{fig:R392} shows calculated  $\Delta\tau$ predicted for the station no 4. As this station is close 
to the event axis (R = 392 m) all predicted $\Delta\tau$ values are shorter than 300 nanosecs.
Three apparent peaks detected at the station 4 are separated by much longer distances
 (see 
Fig.~\ref{fig:MME-4-6}). However, as it has been already mentioned by authors of 
 \cite{peaks-3-8-4-6} the pulses have
a complex shape, what indicates that they consist of more than one signal. As their shapes are too
 complicated 
to  separate individual pulses it is impossible to compare this data with the theoretical predictions.

Summarizing  our attempt to explain these two MME events by means of the strangelet scenario we can 
say:

\begin{itemize}
\item Because of fluctuations it is of course difficult to draw the final conclusion from analysis 
of the individual events.
\item  However, {\bf the features of the analysed individual events do not contradict the assumed
strangelet scenario. They are consistent with such explanation. }
\item  The average  behaviour of other detected MME  events  is in agreement with the proposed model.
 In particular {\bf such MME features as the appareance of 
multipeak structures and 
their dependence on the distance from the detected stations are expected by our strangelet 
 scenario.}
\item Proposed explanation of the MME events does not depend in essential way 
on the values of parameters assumed in the calculations.  
\item   {\bf Further research and simulations are desirable.}
\end{itemize}
   
\section{\Large Summary}

The present work is  continuation of our earlier investigations 
 \cite{9}
concerning the possible connection between the strongly penetrating
cascades
and the strangelets. As it has been shown in our previous studies  the long-range many-maxima 
cascades observed in the lead chambers of the Pamir and Chacaltaya Experiments
could be born during a strangelet passage  through the
apparatus. The data coming from the mountain experiments, however, are not sufficient to decide, if
the exotic cascades  are produced by stable or unstable strangelets. In both cases it was
possible to get by simulations the transition curves resembling the experimental ones.

In this work we checked the hypothesis that the delay signals detected by the Horizon experiments
are produced by the successive strangelet interactions in the atmosphere. Our results indicate that
the extraterrestial strangelets
penetrating deeply into the atmosphere and  degradated in the successive interactions with the air nuclei
could be responsible for the observed multimodal events. Further studies and simulations are required. 

\begin{center}
{\bf Acknowledgments}
\end{center}
I would like to thank Dr Dmitriy Beznosko for sending some papers concerning the unusual events found by the
HORIZON-T experiments and for giving me the permission to use the published there data in my work.
Thank you for careful reading of my manuscript and for many
essential remarks allowing for better understanding of the HORIZON-T experimental results.
 I am especially grateful for the helpful discussions, for pointing out
many problems with the data interpretation and for comments and questions, which hopefully, will help
to solve the mystery of the Chacaltaya, Pamir and Horizon-T exotic events, possibly by means
of the same scenario.

\end{document}